\documentclass{aastex631}

\newcommand{\msun}{M$_{\odot}$ }

\graphicspath{{./}{figures/}}

\begin{document}

\title{High-Resolution, Mid-Infrared Color Temperature Mapping of the Central 10 Arcseconds of the Galaxy}

\author{Cuc K. Dinh}
\affiliation{ Institute for Astronomy, University of Hawai'i, Monoa \\
2680 Woodlawn Drive\\
Honolulu, HI 96822}

\author{Anna Ciurlo}
\affiliation{Department of Physics and Astronomy, University of California, Los Angeles \\
 Los Angeles, CA 90095, USA\\}

 \author{Mark R. Morris}
\affiliation{Department of Physics and Astronomy, University of California, Los Angeles \\
 Los Angeles, CA 90095, USA\\}

\author{Rainer Sch\"{o}del}
\affiliation{Instituto de Astrofísica de Andalucía (CSIC), University of Granada, \\
Glorieta de la astronomía s/n, 18008 Granada, Spain\\}

\author{Andrea Ghez}
\affiliation{Department of Physics and Astronomy, University of California, Los Angeles \\
 Los Angeles, CA 90095, USA\\}

\author{Tuan Do}
\affiliation{Department of Physics and Astronomy, University of California, Los Angeles \\
 Los Angeles, CA 90095, USA\\}

% \collaboration{6}{(AAS Journals Data Editors)}

% \author{Butler Burton}
% \affiliation{Leiden University}
% \affiliation{AAS Journals Associate Editor-in-Chief}

% \author{Amy Hendrickson}
% \altaffiliation{AASTeX v6+ programmer}
% \affiliation{TeXnology Inc.}

% \author{Julie Steffen}
% \affiliation{AAS Director of Publishing}
% \affiliation{American Astronomical Society \\
% 1667 K Street NW, Suite 800 \\
% Washington, DC 20006, USA}

% \author{Magaret Donnelly}
% \affiliation{IOP Publishing, Washington, DC 20005}

%%%%%%%%%%%%%%%%%%%%%%%%%%%%%%%%%%%%%%%%%%%%%%%%%%%%%%%%%%%%%%%%%%%%%%%%%%%%%%%%%
\begin{abstract}
%%%%%%%%%%%%%%%%%%%%%%%%%%%%%%%%%%%%%%%%%%%%%%%%%%%%%%%%%%%%%%%%%%%%%%%%%%%%%%%%%

The neighborhood of the Galactic black hole boasts a plethora of extended interstellar gas and dust features as well as populations of compact (unresolved, or marginally resolved) features such as the G objects. Most are well manifested in the infrared.  To disentangle and characterize the infrared structure of extended features and identify compact sources, we used 3.8~$\mu$m (L' filter) data from the NIRC2 imager at the Keck Observatory and 8.6~$\mu$m (PAH$1$ filter) data from the VISIR imager at the Very Large Telescope (VLT) to produce the highest-resolution mid-IR color-temperature map of the inner half-parsec of the Galactic Center to date. From this map, we compile a catalog of features that stand out from their background. In particular, we identify $33$ compact sources that stand out against the local background temperature, $11$ of which are newly identified and are candidates for being members of the G objects population. Additionally, we resolve and  newly characterize the morphology of several known extended features. These results prepare the way for ongoing and future JWST studies that have access to a greater range of mid-infrared wavelengths, and thus will allow for refined estimation of the trends of dust temperatures.

\end{abstract}
%%%%%%%%%%%%%%%%%%%%%%%%%%%%%%%%%%%%%%%%%%%%%%%%%%%%%%%%%%%%%%%%%%%%%%%%%%%%%%%%%

%%%%%%%%%%%%%%%%%%%%%%%%%%%%%%%%%%%%%%%%%%%%%%%%%%%%%%%%%%%%%%%%%%%%%%%%%%%%%%%%%
\section{Introduction} \label{sec:Introductions}
%%%%%%%%%%%%%%%%%%%%%%%%%%%%%%%%%%%%%%%%%%%%%%%%%%%%%%%%%%%%%%%%%%%%%%%%%%%%%%%%%

The Galactic Center is a hub of activity with numerous phenomena occurring on a wide range of scales. The innermost feature is the  $\sim 4 \times 10^{6}$ \msun black hole manifested observationally as the infrared and radio source Sagittarius~A* (Sgr~A*, \citealt{Ghez2008ApJ...689.1044G, Genzel2010RvMP...82.3121G, Do2019Sci...365..664D, GRAVITY2019A&A...625L..10G}). Within a half-parsec of Sgr~A*  is the young nuclear stellar cluster, a dense population  of young high-mass stars producing strong winds and UV radiation (\citealt{Ghez_2003,Paumard2006ApJ...643.1011P, Bartko2010ApJ...708..834B,LuJR18, vonFellenberg2022ApJ...932L...6V}), intermingled with an older population of main-sequence stars and a distribution of late-type giants (\citealt{Do2009ApJ...703.1323D, Schoedel2020A&A...641A.102S}). Various stellar substructures have also been documented, most notably the stellar clockwise disk (\citealt{Paumard2006ApJ...643.1011P,Gillessen2009ApJ...692.1075G,Lockmann2009MNRAS.394.1841L, Yelda2014ApJ...783..131Y}, and other dynamically coherent structures have been proposed as well \citep{Paumard2006ApJ...643.1011P, Lu2005ApJ...625L..51L, vonFellenberg2022ApJ...932L...6V}.

Streams and clumps of dust and gas also occupy the Galactic Center. Such features include the \textsc{Hii} region Sgr~A West, also known as the Mini-spiral, an apparent three-armed spiral of orbiting, ionized gas streams curled within $\sim$ 2~pc of Sgr~A* (\citealt{ekers1983A&A...122..143E,Lo1983Natur.306..647L, Paumard+04, ZMGA09,Kunneriath2012A&A...538A.127K}), the low-density bubble known as the Mini-Cavity, located at the intersection of the Mini-spiral's Northern Arm and Eastern Arm/Bar (\citealt{Morris+YZ85a,Yusef-Zadeh1989IAUS..136..443Y, Lutz1993ApJ...418..244L,Paumard+04, Schodel2007A&A...469..125S}), and a number of smaller-scale extended features (e.g., \citealt{Yusef_Zadeh_2016ApJ...819...60Y, Yusef_Zadeh_2020MNRAS.499.3909Y} and \citealt{Ciurlo2023ApJ...944..136C}).

The Galactic Center also hosts a number of dust-enshrouded sources (e.g., \citealt{Eckart2004ApJ...602..760E}). Among these, the recently characterized population of features referred to as``G objects'' have been suggested to be the residual of binary stellar mergers \citep[e.g.,][and references therein]{Ciurlo2020Natur.577..337C}. Other works have proposed that they are young stellar objects, \citep{Murray-Clay+Loeb12, Scoville+13, Valencia-S+15, Peissker2023ApJ...944..231P}. 

Young stellar objects have been invoked also to interpret IRS~13N, a cluster of embedded sources that has been suggested  to host extremely young stellar objects about a Myr old (\citealt{Muzic2008A&A...482..173M, Jalali2014MNRAS.444.1205J}).
However, it IRS13~N could also consist of unrelated stars that are passing through a dense dust cloud, heating the local dust in their path. In any case, the presence of young stellar objects in this region is still debated. 

Immediately south of IRS~13N lies IRS~13E, an apparent stellar concentration harboring a few Wolf-Rayet (WR) stars and surrounded by a dense, extended cloud of dusty gas (\citealt{Fritz2010MNRAS.401.1177F,eckart2013A&A...551A..18E}).  It has been suggested that the IRS~13E complex might host an intermediate-mass black hole, which would  account for its apparent cohesion despite the strong tidal forces exerted by the nearby Sgr~A*, which would be expected to disrupt the stellar concentration (\citealt{Maillard2004A&A...423..155M, Tsuboi2019PASJ...71..105T}). 
 That hypothesis could also account for the observed X-ray emission centered on the complex (\citealt{Baganoff+03}). However, this argument has been challenged in favor of a collision of the strong winds of the two WR stars in this cluster, which can account for the X-ray source (\citealt{Zhu2020ApJ...897..135Z,Wang2020MNRAS.492.2481W}).  Furthermore, the notion of IRS13E as a gravitationally bound system has also been questioned based on the proper motion of the different components (\citealt{Schoedel2005ApJ...625L.111S, Fritz+2010b}).  

In Summary, this complex region is characterized by a multitude of sources of varied origins and different components (stars, dust, gas etc.) interacting in a variety of ways.

Here, we probe this rich region through measurements of the color temperature of the dust emission and the associated optical depth. To do so, we compare 3.8~$\mu$m (L') data from the NIRC2 imager at the Keck Observatory and 8.6~$\mu$m (PAH1) data from the VISIR imager at the Very Large Telescope (VLT) and derive the highest resolution dust color temperature map of the inner half-parsec of the Galactic Center. This map allows us to catalog features that stand out from the background: we disentangle and characterize the color temperature structures of extended features and identify compact sources. 

Various mid-IR studies of the central parsec have been conducted over the years (\citealt{Gezari1985ApJ...299.1007G,Cotera+99, Clenet2004, Viehmann2006ApJ...642..861V, Bhat2022ApJ...929..178B}). There has been a number of technological advancements since the last documented color-temperature map of the Galactic Center (\citealt{Gatley1977ApJ...216..277G, Cotera+99,Gezari2003ANS...324..573G}) and higher sensitivity and higher resolution are now available, allowing us to probe the dust temperature and optical depth in unprecedented detail. 

This paper is organized as follows: in Section \ref{sec:Observations}, we describe the observations used for this study, in Section \ref{sec:Analysis} we detail the photometric and astrometric calibration, extinction correction and the derivation of the temperature and optical depth measurements. We present our maps and detail our findings for individual sources in Section\ref{sec:Results}, and raise a general discussion point in Section \ref{sec:disc}. The conclusions of this study are summarized in Section \ref{sec:Summary}.

% [htb]
\begin{figure*}
    \centering
    \includegraphics[width=0.98\textwidth]{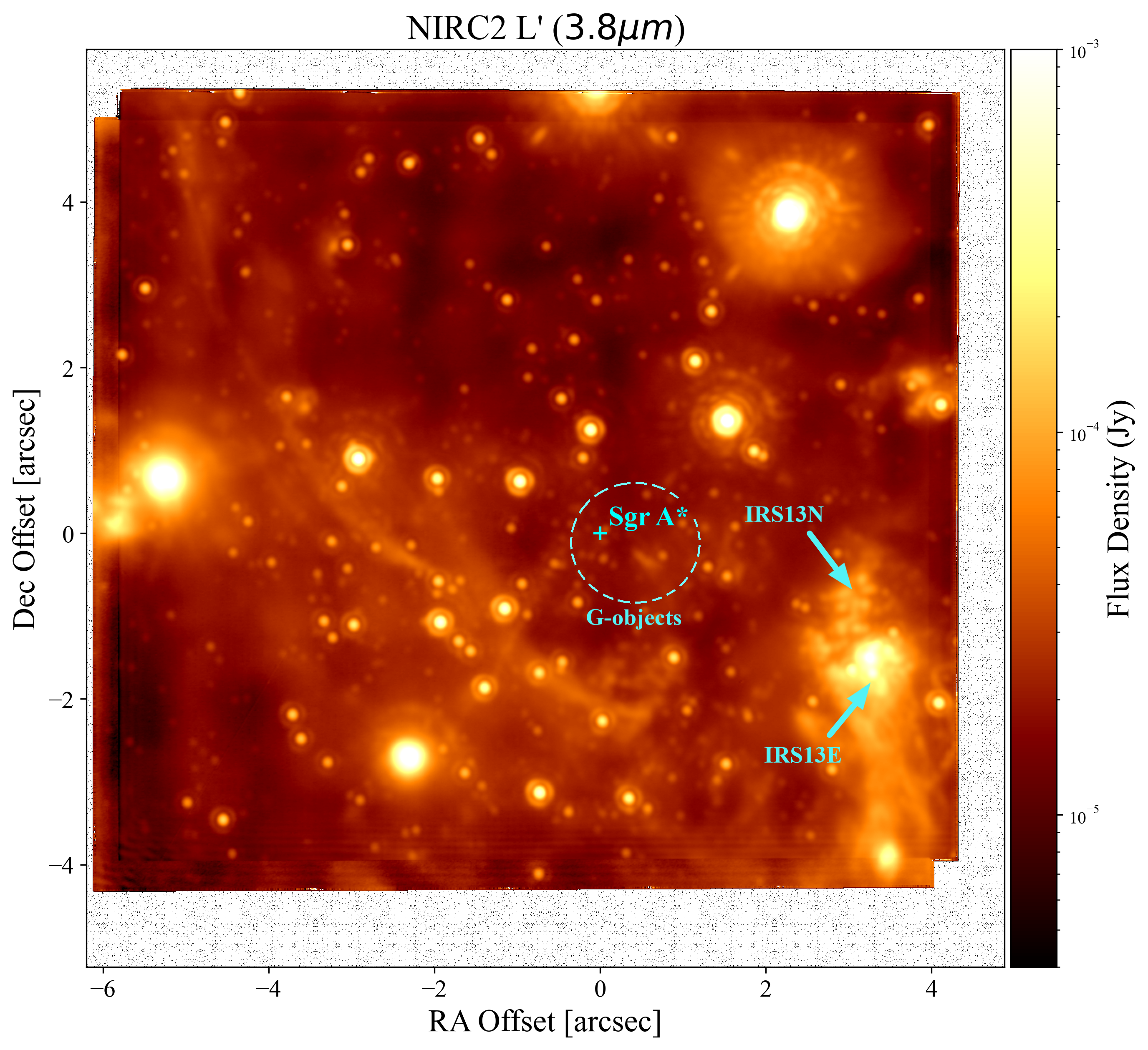}
    \caption{The inner half-parsec region of the Galactic Center as observed in the L' band (3.8~$\mu$m) of the NIRC2 imager at the Keck Observatory. The coordinate offsets in this figure, and in all other figures in this paper, are relative to the position of Sgr~A* \citep{Reid2004ApJ...616..872R}, which is marked with a cyan cross. The cyan circle encloses the domain of the known G-objects (\citealt{Ciurlo2020Natur.577..337C}).}
    \label{fig:Lp_band}
\end{figure*}

%%%%%%%%%%%%%%%%%%%%%%%%%%%%%%%%%%%%%%%%%%%%%%%%%%%%%%%%%%%%%%%%%%%%%%%%%%%%%%%
\section{Observations} \label{sec:Observations}
%%%%%%%%%%%%%%%%%%%%%%%%%%%%%%%%%%%%%%%%%%%%%%%%%%%%%%%%%%%%%%%%%%%%%%%%%%%%%%%%%

Data in the near-infrared (NIR) regime were obtained on July $16$, 2017 through the NIRC2 near-infrared imager of the W. M. Keck Observatory in the L' band (central wavelength of $\lambda_{L'}$=3.8~$\mu$m), with a pixel scale of $0.01$''. The field of view spans 11.08 '' centered on Sgr~A*. The full-width at half-maximum (FWHM) for an isolated star is about $0.086$''. These data were obtained as part of the Galactic Center Orbits Initiative (GCOI\footnote{An ongoing project, presently with a 28-year database obtained using imaging and spectroscopy with the W.M. Keck Observatory. PI: A.M. Ghez}) and have been previously reported in  \cite{Ciurlo+23}. 

Data in the MIR were obtained on the nights of July 15/16th and 25/26th 2017 with the VISIR imager of the Very Large Telescope (VLT) in the PAH$1$ band (central wavelength $\lambda_{PAH1}$=8.6~$\mu$m) and pixel scale 0.045'' (P.I.: R.  Sch\"odel \footnote{Based on observations collected at the European Organisation for Astronomical Research in the Southern Hemisphere under ESO programmes 299.B-5036(B) and 299.B-5036(C).}) A total of 168\,630 individual on-source exposures of 0.021 seconds were obtained with VISIR's so-called burst mode, corresponding to a total on-source time of 3540s.   The FWHM for an isolated source is about $0.191$''.  The observed field of view spans 23.04'' in Right Ascension and 38.70'' in Declination, centered on Sgr~A*. The data were reduced with the speckle-holography described in \cite{Schodel2013MNRAS.429.1367S}.

%%%%%%%%%%%%%%%%%%%%%%%%%%%%%%%%%%%%%%%%%%%%%%%%%%%%%%%%%%%%%%%%%%%%%%%%%%%%%%%%%
\section{Data Analysis } \label{sec:Analysis}
%%%%%%%%%%%%%%%%%%%%%%%%%%%%%%%%%%%%%%%%%%%%%%%%%%%%%%%%%%%%%%%%%%%%%%%%%%%%%%%%%

\subsection{Astrometric and photometric calibration}

We photometrically calibrated the NIR data using 9 reference stars of known magnitude  \citep{Schoedel2010refId0, Gautam2019ApJ...871..103G} \footnote{The 9 stars utilized in photometric corrections for the NIR dataset: S~5-183, S~3-6, S~3-19, S~3-30, S~3-22, S~6-86, S~5-214, S~1-5, and S~3-207}. We used a circular aperture to extract the flux and an annular aperture to correct for the local background. The aperture radius and inner and outer annulus radii were chosen to be respectively $2$, $2.5$, and $3$ times the averaged FWHM of $0.085$'', which was derived by fitting a 2D Gaussian to each reference source and producing an average of the values obtained. 

To correct for the fact that these aperture sizes do not encapsulate the whole source flux, we used  the well-isolated star IRS~33E to compare the flux extracted from the chosen aperture and annulus with the total flux, obtained using aperture and annulus radii of $3.2$, $3.5$, and $4.2$ times the average FWHM, which fully encapsulates the flux for  IRS~33E. This yielded an aperture correction factor of $1.08\pm0.13$. The uncertainty was derived from the dispersion in the width of the fitted Gaussians to the photometric calibrators. 

Comparing the measured flux to the known magnitude of the reference stars, we find a conversion factor of $(8.07 \pm 1.12) \times 10^{-9}$ Jy / $\frac{\text{DN}}{s}$. The uncertainty in the magnitudes of the calibration stars is negligible with respect to the standard deviation of our calibration factor, which leads to an uncertainty of 15.8$\%$ . 

We astrometrically calibrate the NIR data by selecting $29$ well-isolated sources of known absolute position (\citealt{Jia2023arXiv230202040J}), and distributed broadly over the field of view \footnote{The $29$ stars utilized in astrometric corrections for the NIR dataset: IRS~16NE, IRS~16NW, S~2-16, IRS~29N, s~5-183, IRS~33E, IRS~3E, S~4-207, S~2-32, S~1-5, S~1-22, S~2-21, S~3-5, S~3-207, S~5-168, S~3-26, S~3-374, S~3-291, S~4-4, IRS~33N, S~4-2, IRS~7SE, S~5-213, S~3-370, S~1-14, S~3-288, S~4-24, S~3-198, S~4-71.}. We measure the detector position of each source centroid using PSF fitting as has been done in previous publications \citep{Gautam2019ApJ...871..103G}. We compute the transformation between the measured detector positions and the known absolute positions through a  linear least squares fit, assuming the position of Sgr~A* reported by \cite{Reid2004ApJ...616..872R}. 

\begin{figure*}
    \centering
    \includegraphics[width=0.98\textwidth]{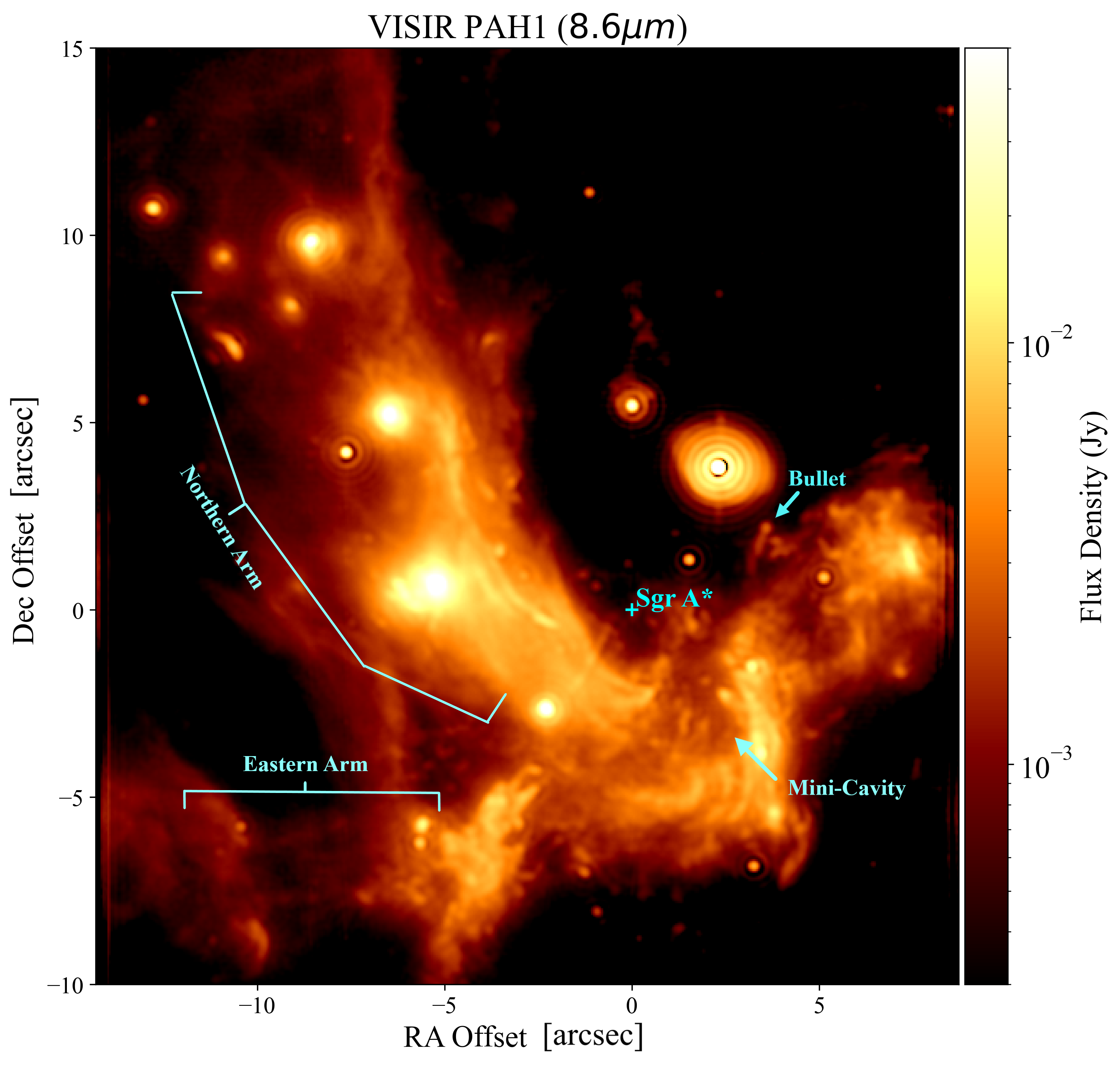}
    \caption{The inner parsec region of the Galactic Center as observed with the PAH$1$ band (8.6~$\mu$m) of the VISIR imager at the Very Large Telescope. The position of Sgr~A* is marked with a cyan cross. Extended features such as the Northern Arm curling in along the left-hand side of the image dominate this band.}
    \label{fig:PAH$1$_band}
\end{figure*}

 The photometric calibration of the 8.6~$\mu$m data was performed with standard star observations. Five IRS stars were used as a PSF reference with iterative fitting, and background determination was included to account for diffuse emission. 

The MIR data were astrometrically calibrated using IR-bright stars of known position that are detected in the radio as SiO masers (\citealt{Reid2004ApJ...616..872R, Sakai2019ApJ...873...65S, Darling2023ApJ...955..117D}), obtaining an absolute astrometric accuracy better than $0.01$ arcsecond. 

\subsection{Extinction correction} \label{sec:Extinction}

 There are local variations of extinction in the Galactic Center of up to 1 magnitude across the field of view in the K-band \citep{Schoedel2010refId0}. For this reason we verified the impact of spatially varying extinction in order to disentangle real emission variations from variations due to differences in the foreground extinction. To do so we considered two extinction maps: star-derived extinction (Ks-band, \citealt{Schoedel2010refId0}) and gas-derived extinction (V-band, \citealt{Scoville2003ApJ...594..294S}). We considered both maps to account for the fact that interstellar medium and stellar features might not be coincident along the line of sight, and hence might be subject to different extinction variations. We extrapolated A$_{3.8}$, A$_{8.6}$ and A$_{K_{s}}$ using the results presented in \cite{Fritz2011ApJ...737...73F}. 
 
The intrinsic flux $\text{F}_{0}$, observed flux $\text{F}$, and extinction A($\lambda$) are related such that:

\begin{equation}
    A(\lambda)= -2.5\log(\frac{F_{observed}}{F_{expected}})
\end{equation}

Given the flux measurements in two wavelengths ($\lambda_{3.9}$ and $\lambda_{8.6}$) , the corresponding extinctions propagate such that the the observed flux ratio is:

\begin{equation}
    \frac{F_{3.8}}{F_{\text{8.6}}} = 10^{\frac{A(\lambda_{\text{8.6}})-A(\lambda_{3.8})}{2.5}}(\frac{F_{3.8,0}}{F_{\text{8.6},0}})
\end{equation}

Using the extinction map from (\citealt{Schoedel2010refId0}), which reports the extinction at Ks, we find that in order to correct our observations we need to compute the corresponding extinction at 3.8~$\mu$m:

\begin{equation}
    \frac{F_{3.8}}{F_{8.6}} = 10^{\frac{(A(\lambda_{8.6})-A(\lambda_{3.8}))A'(\lambda_{K_{s}})}{2.5A(\lambda_{K_{s}})}}(\frac{F_{3.8,0}}{F_{8.6,0}})
\end{equation}

The same method applies for the extinction map from \cite{Scoville2003ApJ...594..294S}.

We find that, although extinction dims the observed flux by a factor of 0.34 at L' and of 0.19 at 8.6~$\mu$m, there is no significant spatial variation in the dust temperature map  because the dependence on wavelength does not vary much between those two wavelengths. Thus, hereafter all our results are presented without extinction correction.

\subsection{Dust Color Temperature and Optical Depth Measurements} \label{sec:Temperature}

The color temperature of a source can be determined by comparing its observed radiance at two different wavelengths. We first convolved the L' dataset with the point-spread function (PSF) of the 8.6 $\mu$m data, which has a much larger FWHM than the L' data, in order to match the resolution between two datasets. Then, pixel by pixel, we pair the spatial locations between both datasets through WCS transformation functions from the astropy python library. This way we obtained a map of the ratio of flux densities between L' and PAH$1$ ($\frac{3.8}{8.6}$). 

Finally, we find the color temperature by numerically solving for Planck function:At thermal equilibrium, the color temperature (T$_{c}$) of an object can be found by considering the ratio of its radiance at two different wavelengths ($\lambda_{3.8}$ and $\lambda_{8.6}$). Furthermore, The radiance ratio can be swapped with the flux density ratio ($\frac{F_{3.8}}{F_{8.6}}$) if the unit solid angle is the same. Thus, we solve for the color temperature, T, using:

\begin{equation}
    \frac{F_{3.8}}{F_{8.6}}=(\frac{\lambda_{8.6}}{\lambda_{3.8}})^{5}\frac{e^{\frac{hc}{\lambda_{8.6}k_{b}T}}-1}{e^{\frac{hc}{\lambda_{3.8}k_{b}T_{c}}}-1}
\end{equation}

We also calculate the corresponding optical depth at 3.8 $\mu$m by comparing the observed flux, $F_{\nu}$, and its corresponding blackbody irradiance, $B_{\nu}(T_{c})$, given the measured temperature:

\begin{equation}
    \tau_{\nu} = - \ln[1 - \frac{4F_{\nu}}{\pi B_{\nu}(T_{c})\text(FWHM)^{2}}]
\end{equation}

%%%%%%%%%%%%%%%%%%%%%%%%%%%%%%%%%%%%%%%%%%%%%%%%%%%%%%%%%%%%%%%%%%%%%%%%%%%%%%%%%
\section{Results}\label{sec:Results}

\begin{figure*}[h]
    \centering
    \includegraphics[width=0.98\linewidth]{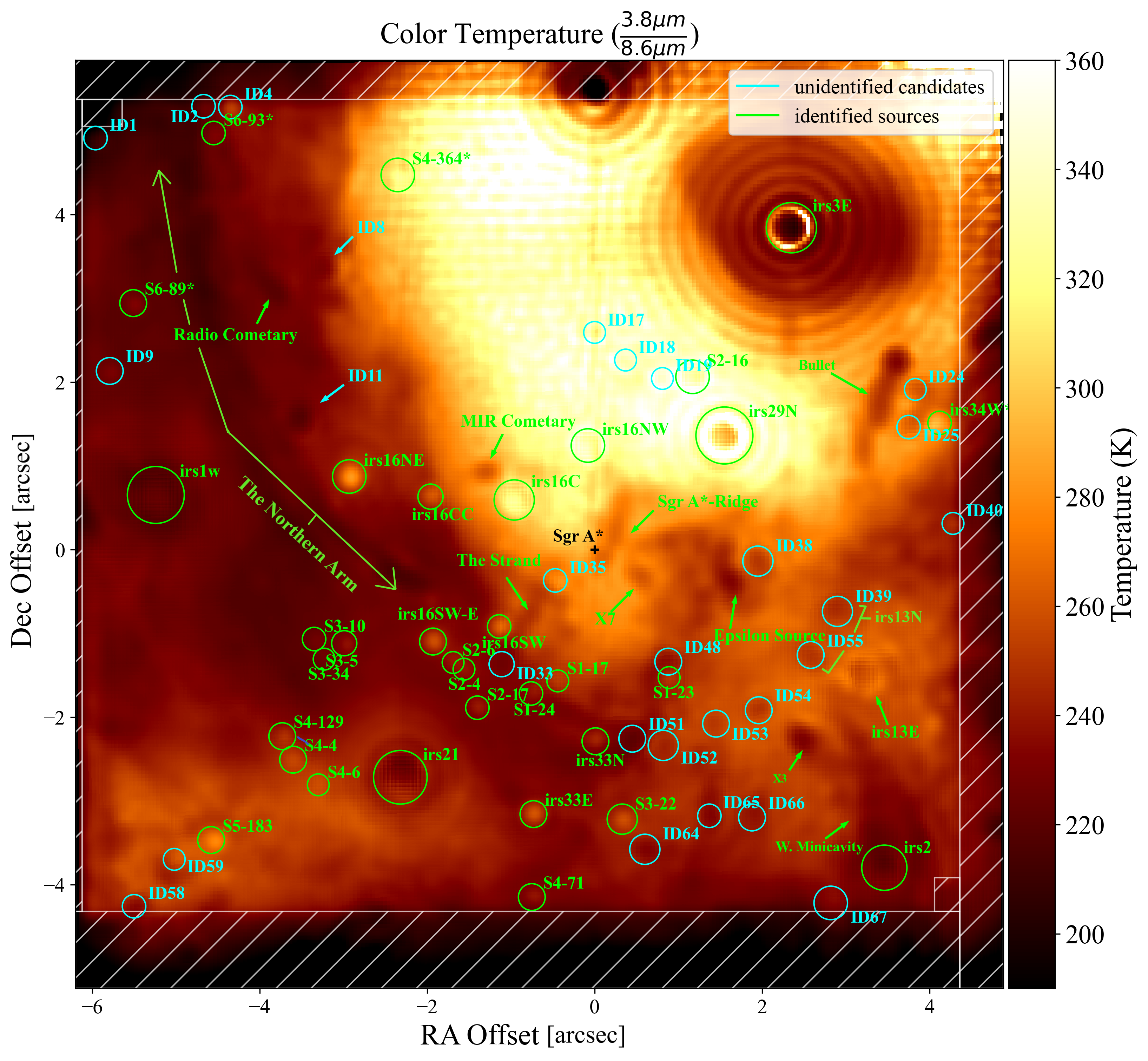}
    \caption{Color temperature map, annotated with all compact objects and extended features. Previously known features are marked in green, newly reported ones are marked in cyan.}
    \label{fig:color_temp}
\end{figure*}

The color temperature map we obtain is presented in Figure~\ref{fig:color_temp}. The map can be divided into three broad regions: the relatively cool Sgr~A West (Mini-spiral) region, curving from the northeast corner downward to the west, the higher temperature region in the northwest, and the bottom region extending from IRS~21 to the western edge of the mini-cavity, having an intermediate color temperature. The fish-shaped dust ridge adjacent in projection to Sgr~A* has previously been referred to as the Sgr~A*-Ridge (\citealt{Stolovy1996ApJ...470L..45S,Schodel2007A&A...469..125S, Schoedel2010refId0}). We report the fluxes and temperatures of several notable sources (see Table~\ref{tab:VALUES} in Appendix \ref{sec:sources}).

\begin{figure*}[h]
    \centering
    \includegraphics[width=0.98\linewidth]{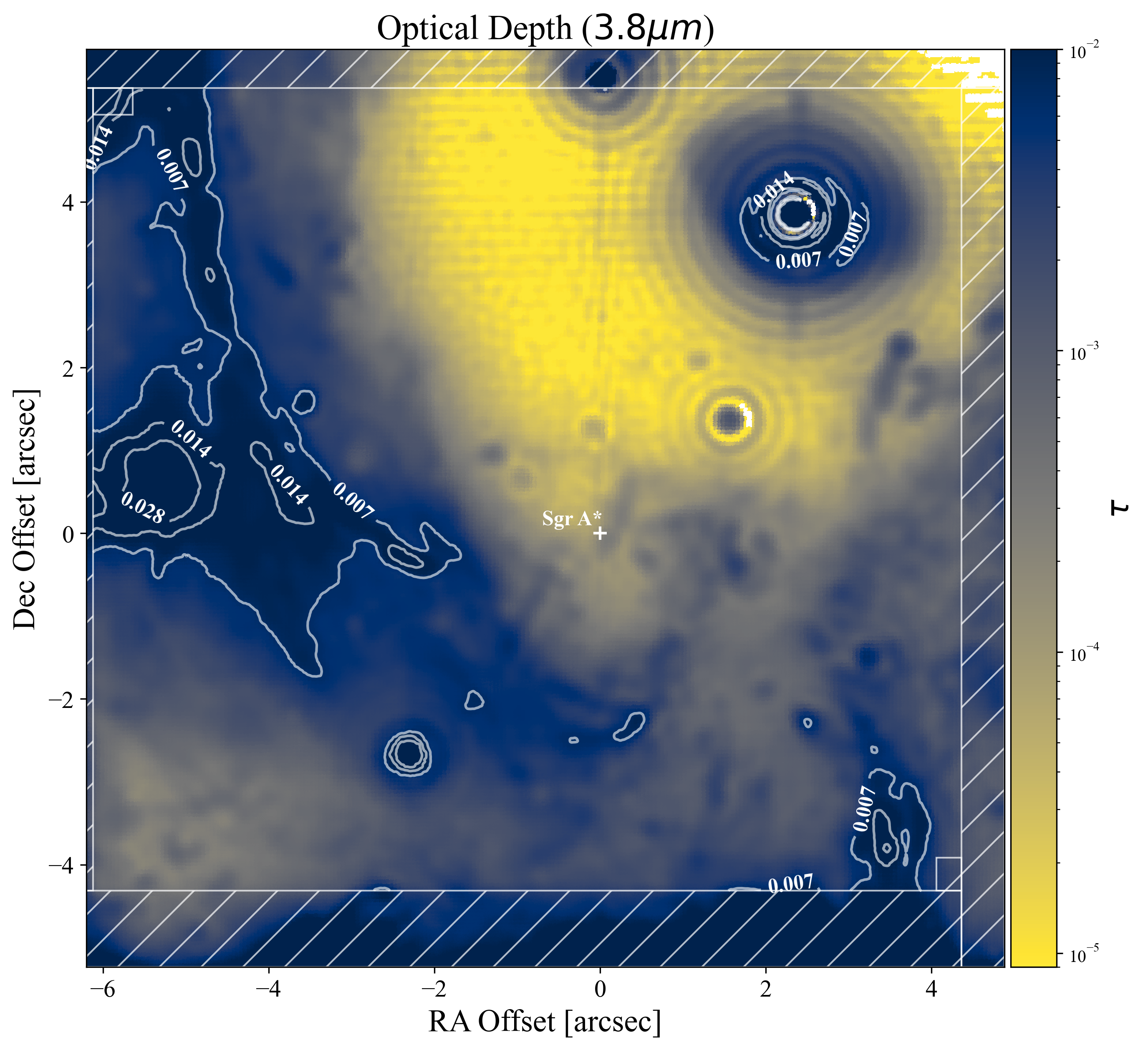}
    \caption{Optical Depth map at L' ($3.8 \mu$m), with contours at $\tau =$0.007, 0.014, and 0.028. The little variation in temperature yields what is essentially an inverted map; low temperature features such as the Northern Arm translate into a higher optical depth analogue. }
    \label{fig:opdep}
\end{figure*}

A corresponding 3.8~$\mu$m optical depth map is presented in Figure~\ref{fig:opdep}. This map mostly appears as an inversion of the color temperature map because of the limited range of the color temperature variations in the region (spanning $\sim$200~K), with lower temperature signatures usually coinciding with higher optical depths. We note that the optical depth is always small enough that smaller color temperatures cannot be ascribed to higher optical depths; that is, regions of smaller color temperature can't simply be explained by foreground absorption of a warmer background.

\subsection{Compact Sources}\label{sec:hot and cold}

Several point-like features stand out in our temperature map. We can qualify them as "hot" or "cool" depending on the temperature difference with respect to their local background.  
We define as hot features those with temperatures 10~K higher than their local background, and as cool features those with temperatures  10~K lower than their background. The source color temperature is computed as the median value within a circular aperture having a radius chosen individually for each source, while the local background temperature is defined as the median value within an annulus of inner and outer radii 1.3 and 2 times the circular aperture radius. The apertures were centered on the source center-of-mass centroids measured with the photutils package in python (\citealt{larry_bradley_2023_7946442}).  In the measurement of the background we take care to remove any overlapping source from the chosen annulus. 
This method resulted in the identification of 33 hot/cold compact sources. Of this sample, 22 are known stars (\citealt{Yelda2014ApJ...783..131Y,Gautam2019ApJ...871..103G}) and 11 are newly reported. We note that many of the identified sources are located to the south and southwest of Sgr~A* (see Fig.\ref{fig:color_temp}), most likely because this area is less obfuscated by nearby extended dust features. 

 With the exception of ID~2 and ID~4, which are at the edge of our field, all hot compact features identified are associated with known stars that are most likely heating surrounding dust in which the star is embedded or very nearby. Thus, we focus on the cool features (\citealt{Eckart2014IAUS..303..269E, Sitarski2016PhDT.......349S}). 
 
\begin{figure}
    \centering
    \includegraphics[width=0.98\textwidth]{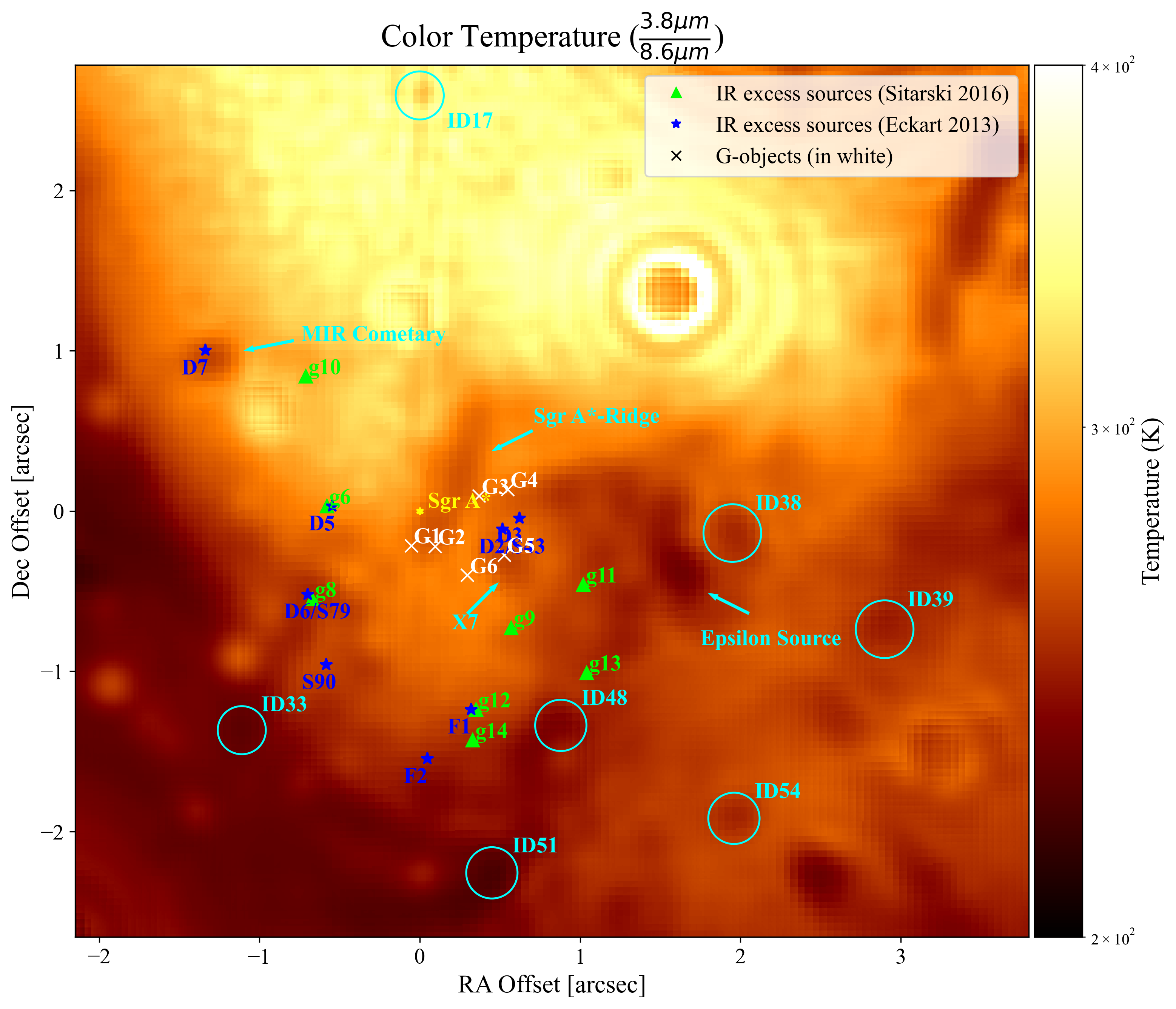}
    \caption{Locations of currently known Infrared Excess sources and G objects,  compared to the identified cold sources within the vicinity. Cool sources ID5~4, ~55, and ~65 are a located beyond this region. The so-called Sgr~A*-ridge superimposed nearby Sgr~A*on the line of sight and not actually within its vicinity in 3D (but see also\citealt{Peissker2020ApJ...897...28P}). }
    \label{fig:IR_excess}
\end{figure}

 As seen in Figure \ref{fig:IR_excess}, with the exception of D$7$ \citep{eckart2013A&A...551A..18E} and the so-called 'MIR Cometary' \citep{Yusef_Zadeh_2016ApJ...819...60Y}, all of the cold compact features we identify are newly reported.  A particularly interesting group amongst the known cool compact sources in this region is the G objects population (\citealt{Ciurlo2020Natur.577..337C}). These objects have been proposed to host a hidden star  enshrouded in dust (although the process that produced this configuration is debated  \citealt{Witzel2014ApJ...796L...8W,Ciurlo2020Natur.577..337C, Peissker2021ApJ...923...69P}), and thus are expected to have cooler thermal signatures than emission-line stars. The cool point-like features we detect are therefore new candidate members of the G objects population. 
To confirm membership in this population, multiple epochs of high-resolution follow-up observations are required. If these are indeed members of the G objects population, that would significantly increase the sample size and radial distribution of these objects and hence push forward the investigation of their nature and origin.
Most of the previously known cool sources (G-objects and/or IR-excess sources) are not detected in our color temperature map likely because they are concentrated in the innermost region where, given the resolution of our data, we are limited by confusion due to extreme crowding. 

\subsection{Extended Features}

Of the many extended gas and dust features (some being the object of past and on-going investigations in this region, see for e.g. \citealt{Yusef-Zadeh1989IAUS..136..443Y,Yusef_Zadeh_2016ApJ...819...60Y}), we examine the thermal structure of the ones showing the most notable features in our color-temperature map. 

\subsubsection{The Bullet}

\begin{figure}
    \centering
    \includegraphics[width=0.98\textwidth]{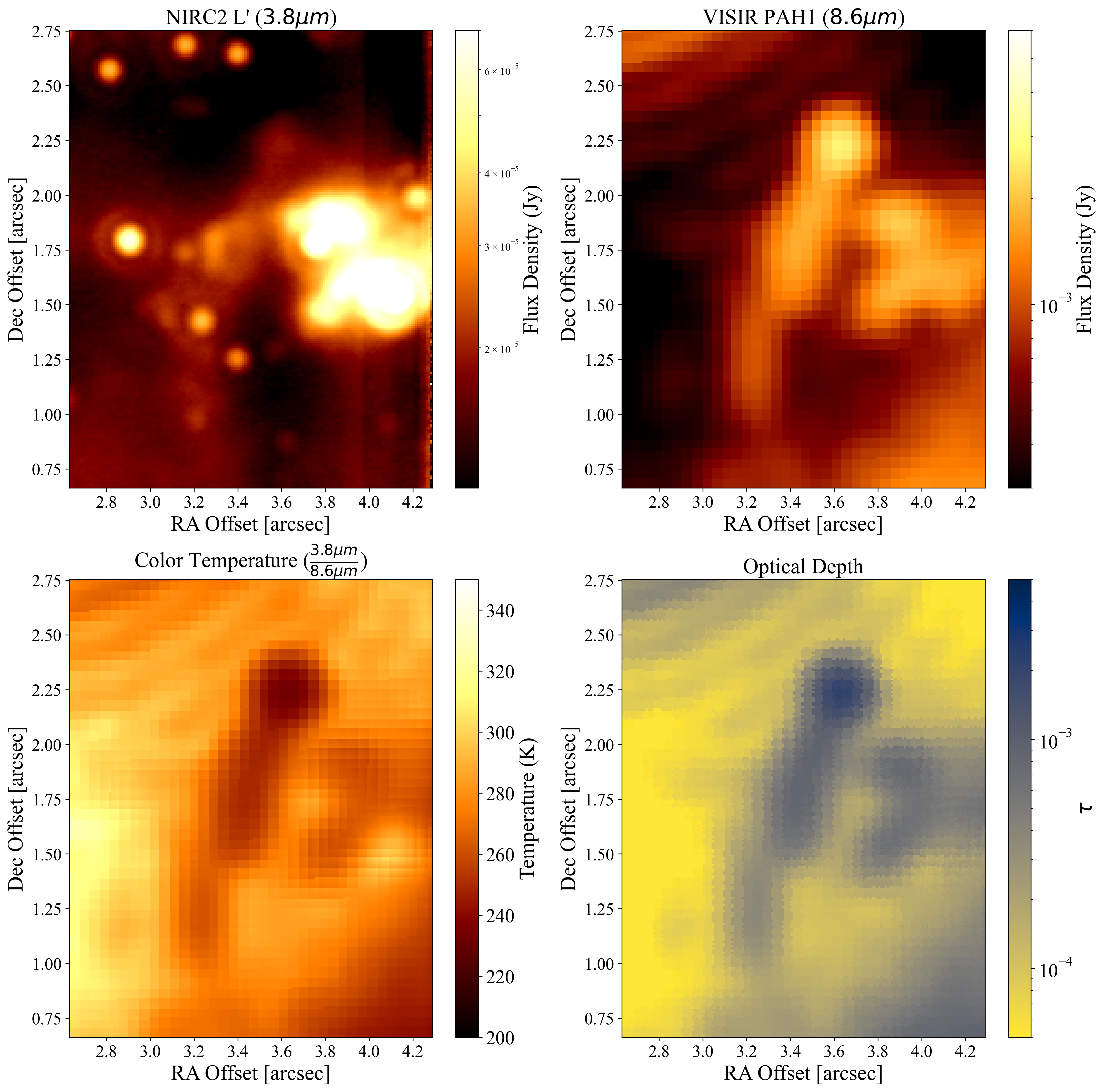}\caption{Comparison of the Bullet feature as observed in the 3.8~$\mu$m band (upper left) , 8.6~$\mu$m band (upper right), color temperature map (lower left), and optical depth map at 3.8~$\mu$m (lower right). We find that the color temperature of the Bullet is lowest at the head, and increases in temperature with increasing distance along its tail. The proper motion of the Bullet shows it to be moving parallel to its length with the head at the leading edge (\citealt{ZhaoGoss98}). }
    \label{fig:bullet_comp}
\end{figure}

The Bullet is a rapidly moving radio feature with a head-tail morphology  \citep{YZ+98,ZhaoGoss98,ZMGA09} that lies to the south of IRS~3E (Figure~\ref{fig:bullet_comp}). In our color temperature map, the compact head of the bullet is about 50~K colder than the local background. The head is trailed by a longer tail that progressively approaches the background temperature with increasing distance from the head. The optical depth gradient along the ridge of the bullet is inversely correlated with the temperature, with the head having the largest optical depth. 
The Bullet is currently not known to be associated with any known star and its origin is undetermined, but its rough morphological similarity to the linear feature X7 raises the possibility that the Bullet and X7 were formed by a similar mechanism, possibly a stellar collision \citep{Ciurlo+23}\footnote{The shape of the Bullet is also similar to that of some pulsar wind nebulae, but their non-thermal emission is not typically accompanied by thermal infrared emission}. 

\begin{figure*}
        \centering
        \includegraphics[width=.45\linewidth]{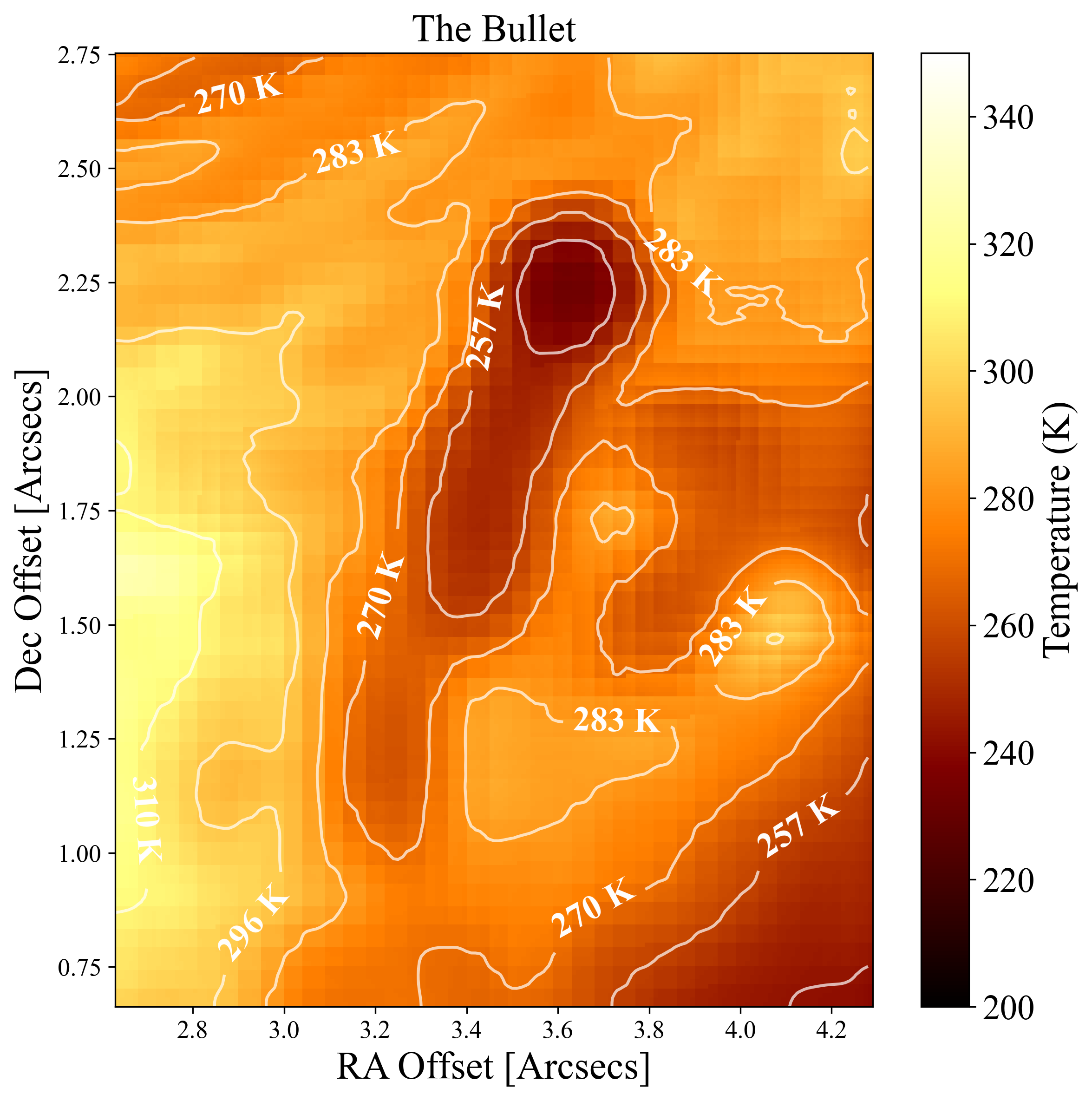}
        \centering
        \includegraphics[width=.5\linewidth]{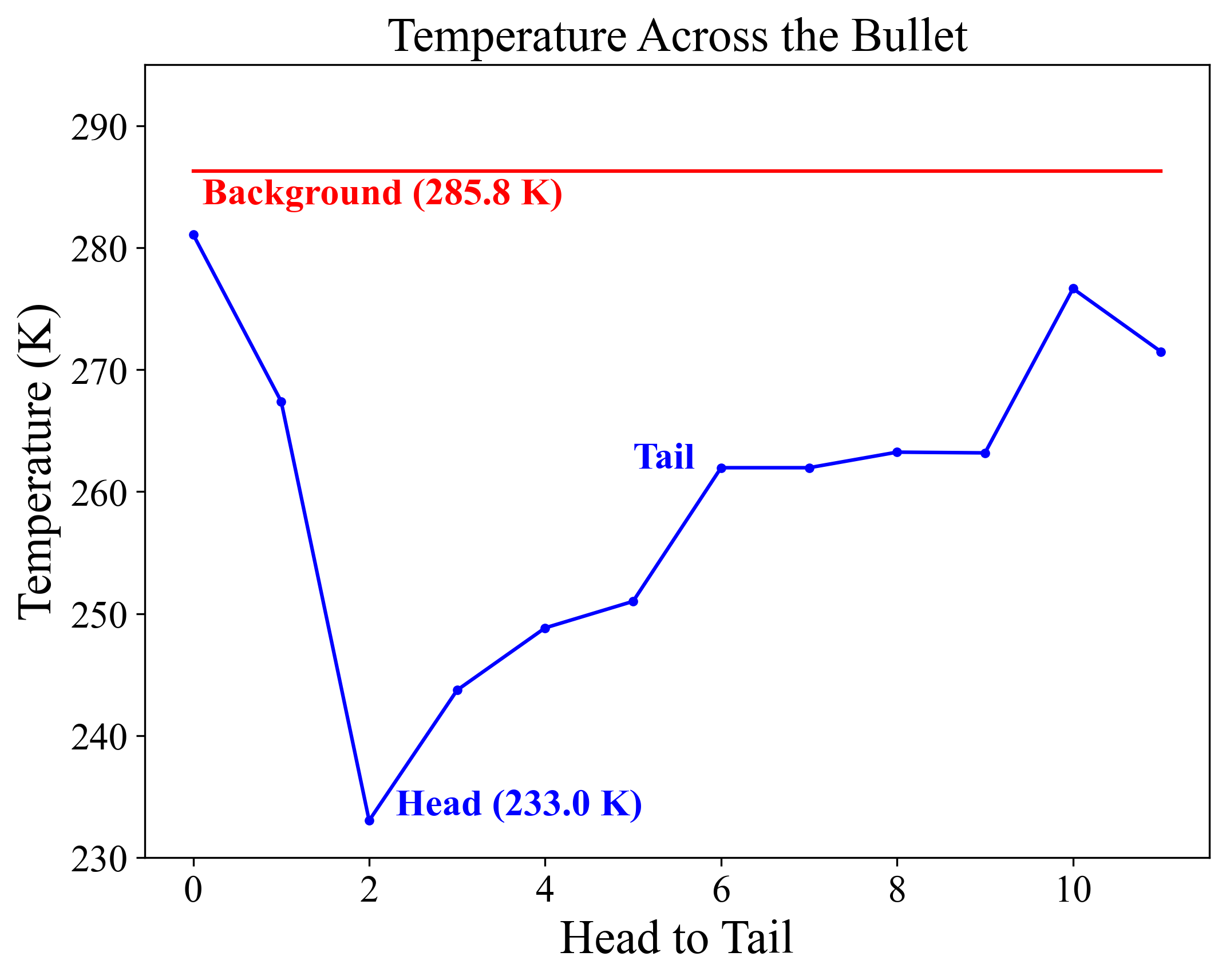}
    \caption{Left: Color temperature contours of the Bullet. Right: color temperature distribution along the Bullet structure. The Head is the reddest part of the bullet at 49.6~K cooler than the local background.}
    \label{fig:bullet comparison}
\end{figure*}

\subsubsection{IRS~13E}

The IRS~13E complex is comprised of various stellar objects that appear clustered in projection, as well as an additional extended feature, IRS~13E3 (hereafter E3).  E3 is nestled in the center of the cluster and has been interpreted as a shock resulting from colliding winds from two luminous WR stars in the complex, E2 and E4 (\citealt{Zhu2020ApJ...897..135Z, Wang2020MNRAS.492.2481W}). X-Ray emission from E3 had previously led some to consider that it could be associated with an intermediate-mass black hole responsible for binding the IRS~13E cluster  \citep{Maillard2004A&A...423..155M, Tsuboi+17_IRS13, Tsuboi+19b}, but only a few of the stars in this cluster share a common proper motion, so others are likely fortuitously close in projection \citep{Schoedel2005ApJ...625L.111S,Fritz+2010b, Zhu2020ApJ...897..135Z}. 

\begin{figure}
    \centering
    \includegraphics[width=0.98\textwidth]{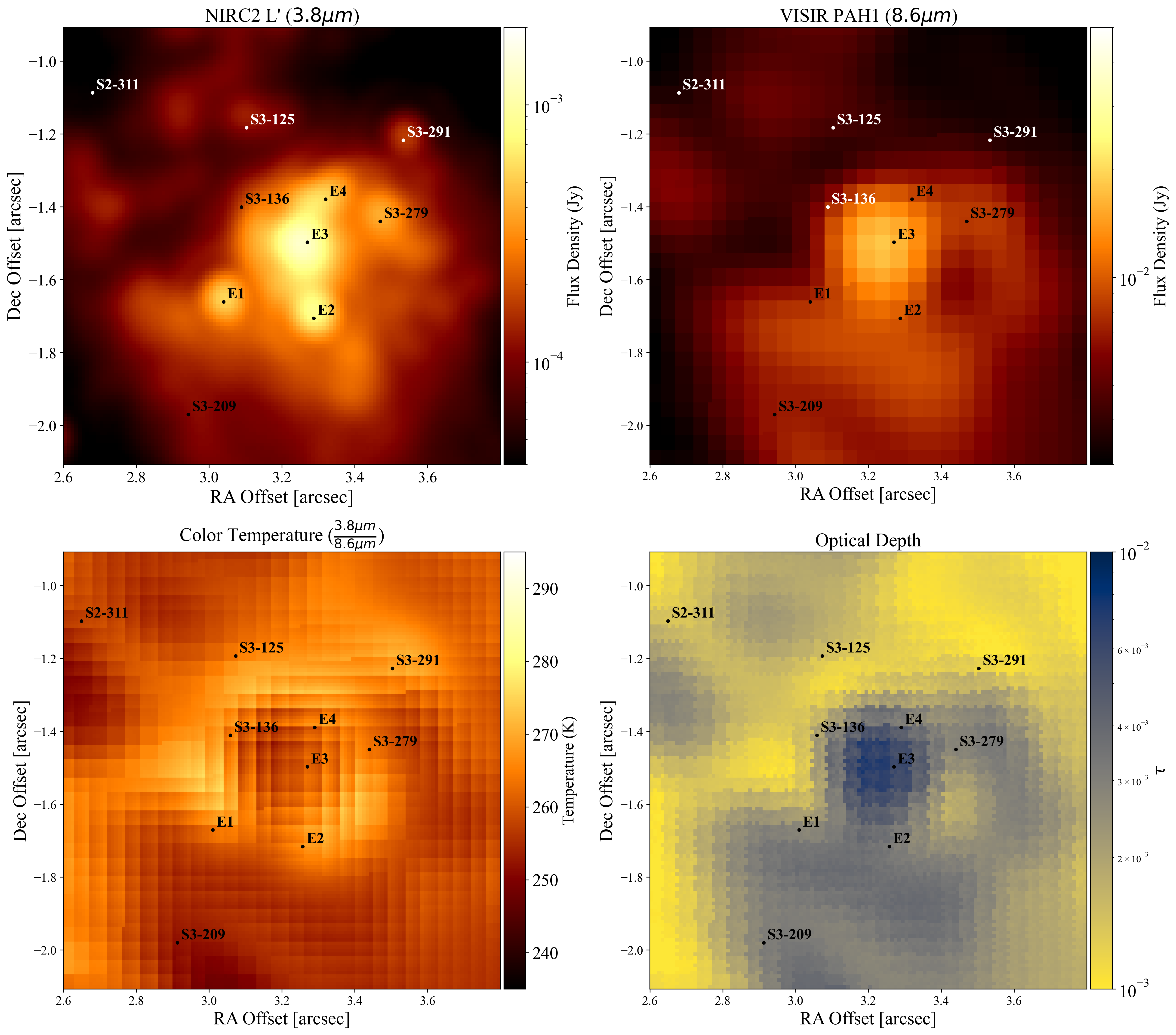}
    \caption{Comparison of the IRS~13E complex as observed in the 3.6~$\mu$m band (upper left) , 8.6~$\mu$m band (upper right), color temperature map (lower left), and 3.8~$\mu$m optical depth map (lower right). The components of the complex have been annotated in the L' band image. The color temperature map shows a warm ring or shell surrounding a colder core.}
    \label{fig:IRS13E_comp}
\end{figure}

 In the temperature map (a zoom-in is shown in the rightmost panel of Figure \ref{fig:IRS13E_comp}) we find that IRS~13E has a cold core compared to the background and is surrounded by a warm shell about 15~K hotter than the core. This roughly circular shell is divided into three warm arcs in the color temperature map. E3, which is somewhat extended at 3.8~$\mu$m (\citealt{Zhu2020ApJ...897..135Z}), is not exactly centered on the color temperature minimum, but its close positional correspondence suggests that it is associated with it. 

 The hypothetical wind collision generating the shock in E3 would probably lead to dust production, which would imply a higher optical depth in the core. The ambient dust surrounding the luminous WR stars is expected to be relatively warmer due to the high radiation density in the region around the WR stars. However, the ring does not bear an obvious spatial relationship to the WR stars, E2 and E4, so this hypothesis might require a particular non-uniform initial distribution of the dust, as well as an accounting for the relative velocity of the WR stars and the ambient dust.

\subsubsection{IRS~13N}

About half an arcsecond north of IRS~13E is a complex of several reddened sources known collectively as IRS~13N \citep{Eckart2004ApJ...602..760E, Muzic2008A&A...482..173M, eckart2013A&A...551A..18E}.  Hydrodynamical simulations were conducted to examine the hypothesis that these objects are young stellar objects (YSOs) formed from compressed molecular clumps in the vicinity of Sgr~A* (\citealt{Jalali2014MNRAS.444.1205J}).  
More recently \cite{Peißker_2023} suggested IRS~13E is also part of this complex and that the whole cluster has recently formed and contains abundant YSOs.

\begin{figure}
    \centering
    \includegraphics[width=0.98\textwidth]{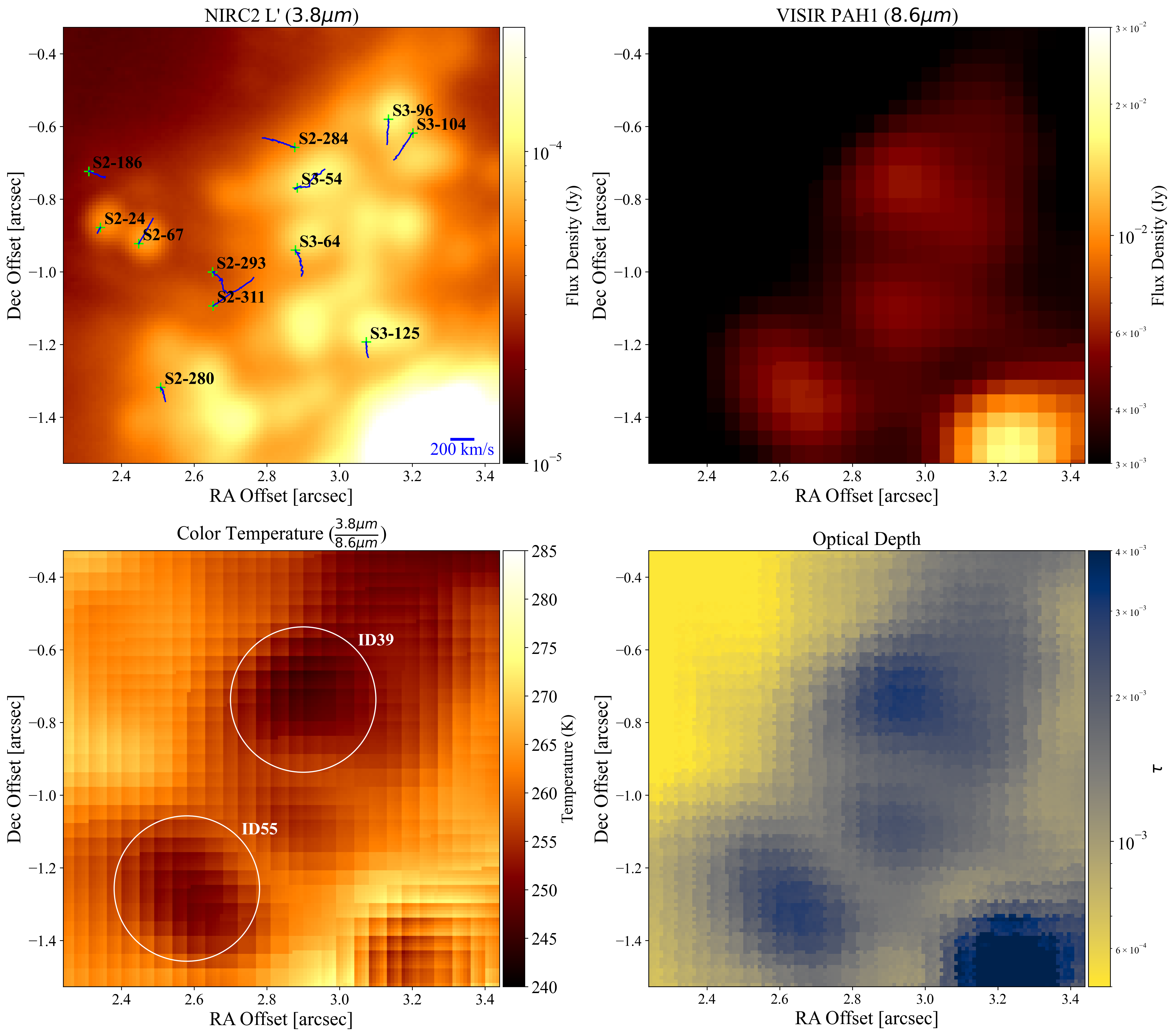}
    \caption{Comparison of the IRS~13N complex as observed at 3.6~$\mu$m (upper left) , 8.6~$\mu$m band (upper right), color temperature (lower left), and optical depth at 3.8~$\mu$m (lower right). Known K-band stars and their proper motion are marked in the L' map \citep{Jia_2019}. Cool features ID55 and ID39 have been marked in the PAH map.}
    \label{fig:IRS1N}
\end{figure}

The IRS~13N complex appears as a "cluster of grapes" in the L' band (upper left panel of Figure~\ref{fig:IRS1N}): several emission peaks, presumably stars embedded in a dense, dusty medium,  are connected by a central "stem" bisecting the complex. The structure in the 8.6~$\mu$m band (and by consequence in the color temperature map, center and right panels of Figure~Fig. \ref{fig:IRS1N}) is less discernible because of the lower spatial resolution. We identify two broad, cool features in our color temperature map: ID55 (South) and ID39 (North). ID55 partially overlaps with a separate emission peak located to the southeast of the IRS~13N complex in the L' image, whereas ID39, as well as the small clump in between ID55 and ID39, appear to constitute the main portion of the complex.

The features in this complex are characterized by a lower color temperature than observed in the ambient background. Because dust-embedded stars are expected to heat their surrounding dust, this could be explained by the dust dominating the background being substantially hotter than the dust around the stars. The peaks in the L' and color temperature maps also do not appear to be coincident, which suggests that the dust around the stars is heated anisotropically. This could either be explained by an inhomogenous dust distribution or by anisotropic irradiance of dust in the region by the light emerging from accretion disks. 

While IRS~13N has been suggested to be a coherent cluster of YSOs, the proper motions of its putative components are inconsistent with this being a bound system (see Figure~\ref{fig:IRS1N} in this paper and Figure~4 in \citealt{Peissker2023ApJ...944..231P}). A viable alternative explanation is that these stars just happen to be passing through a dense clump of dust along their individual trajectories and they are locally heating the dust that immediately surrounds them. 

\subsubsection{ID8 and ID11}

Projected toward the Northern Arm of Sgr~A West/Mini-spiral are two compact features having lower temperature than their immediate surroundings: ID8 and ID11. While these features are not cool enough to meet our 'cool object' criterion, they nonetheless stand out for their interesting morphology.  

\begin{figure}
    \centering
    \includegraphics[width=0.98\textwidth]{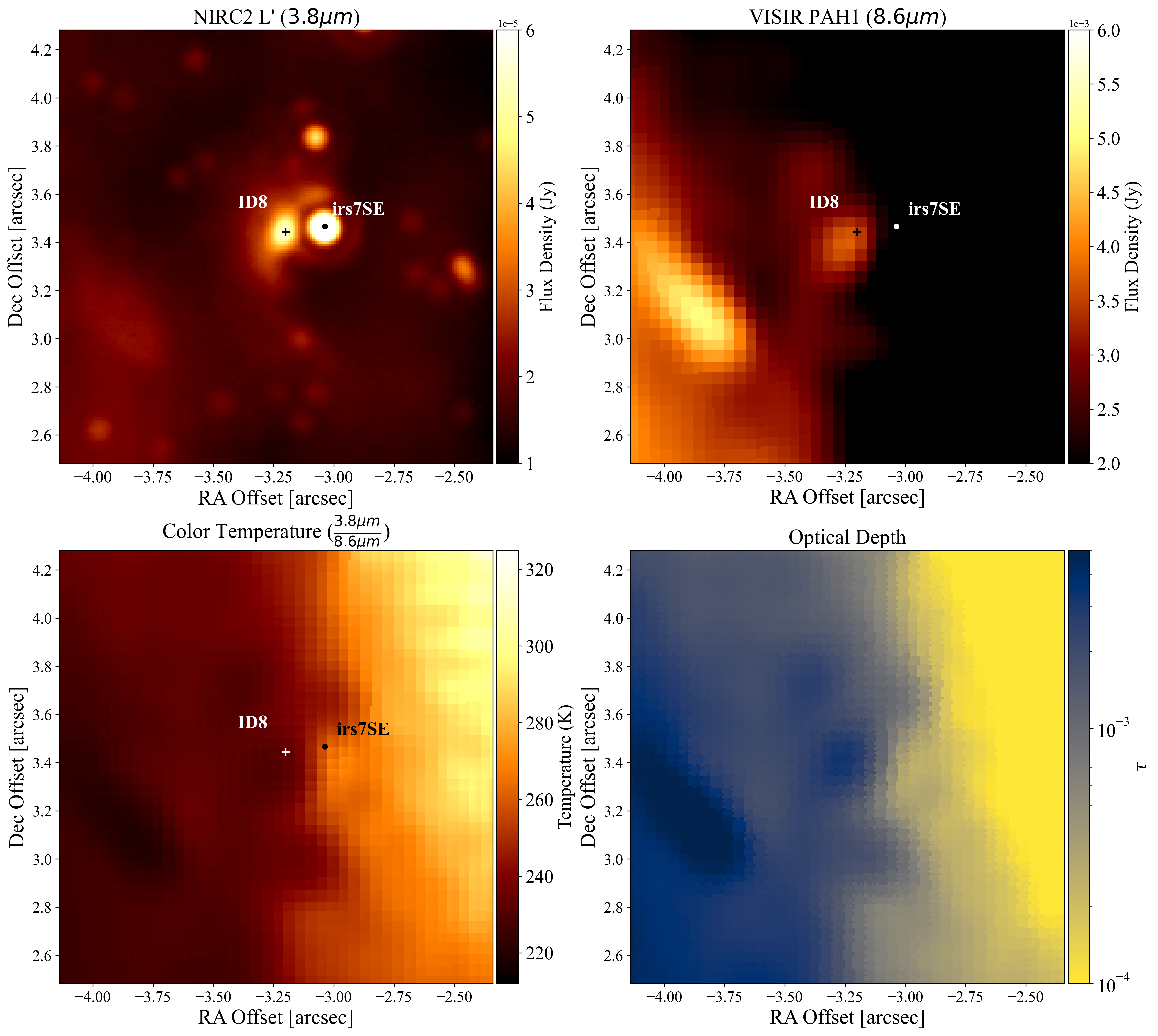}
    \caption{Comparison of ID8 as observed in the 3.6~$\mu$m band (upper left) , 8.6~$\mu$m band (upper right), color temperature map (lower left), and optical depth map at 3.8~$\mu$m (lower right). This candidate corresponds to the shock associated with IRS~7SE. The color temperature map (right) registers a cooler object, beside which is an illuminated portion facing outward of the Northern Arm. } 
    \label{fig:ID8}
\end{figure}

ID8 appears to be centered on a shocked region (\citealt{Yusef_Zadeh_2016ApJ...819...60Y}) flanking the eastern side of  IRS~7SE (see Fig. \ref{fig:ID8}), a WC9 star with a high-velocity outflow (\citealt{Marins2007A&A...468..233M}). ID8 and the nearby star IRS~7SE are superimposed on a sharp color temperature gradient that marks the edge of the Northern Arm, likely resulting from the interaction of the dust and gas in the Northern Arm with the radiation and winds from the young nuclear cluster. ID8 is the peak of a clump of dust visible in the optical depth map. 

\begin{figure}
    \centering
    \includegraphics[width=0.98\textwidth]{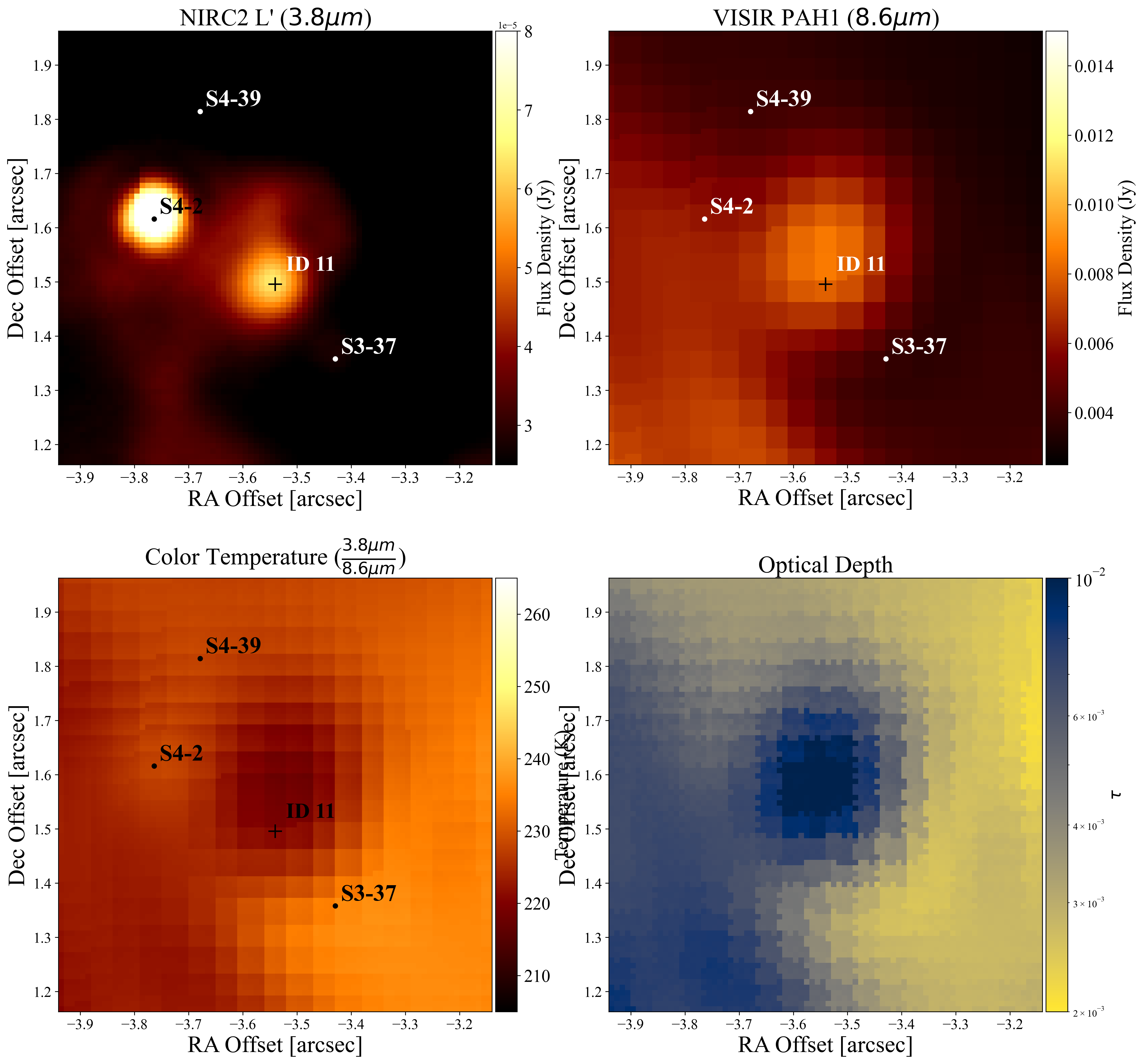}
    \caption{Comparison of ID11 as observed in the  3.6~$\mu$m band (upper left) , 8.6~$\mu$m band (upper right), color temperature map (lower left), and optical depth map at 3.8~$\mu$m (lower right). Note the the cooler upwards stretch in the color temperature map (right).}
    \label{fig:ID11}
\end{figure}

ID11 is located further South (Fig.~\ref{fig:ID11}), between the stars S4-2 and S3-37  and lies immediately to the northwest of the bright ridge of the Northern Arm  (Figure.~ \ref{fig:color_temp}). The small emission feature superimposed  on ID11 and the region immediately North of it appear to be cooler and to have a higher column density than the local background. 

\subsubsection{The Strand}

\begin{figure}
    \centering
    \includegraphics[width=0.98\textwidth]{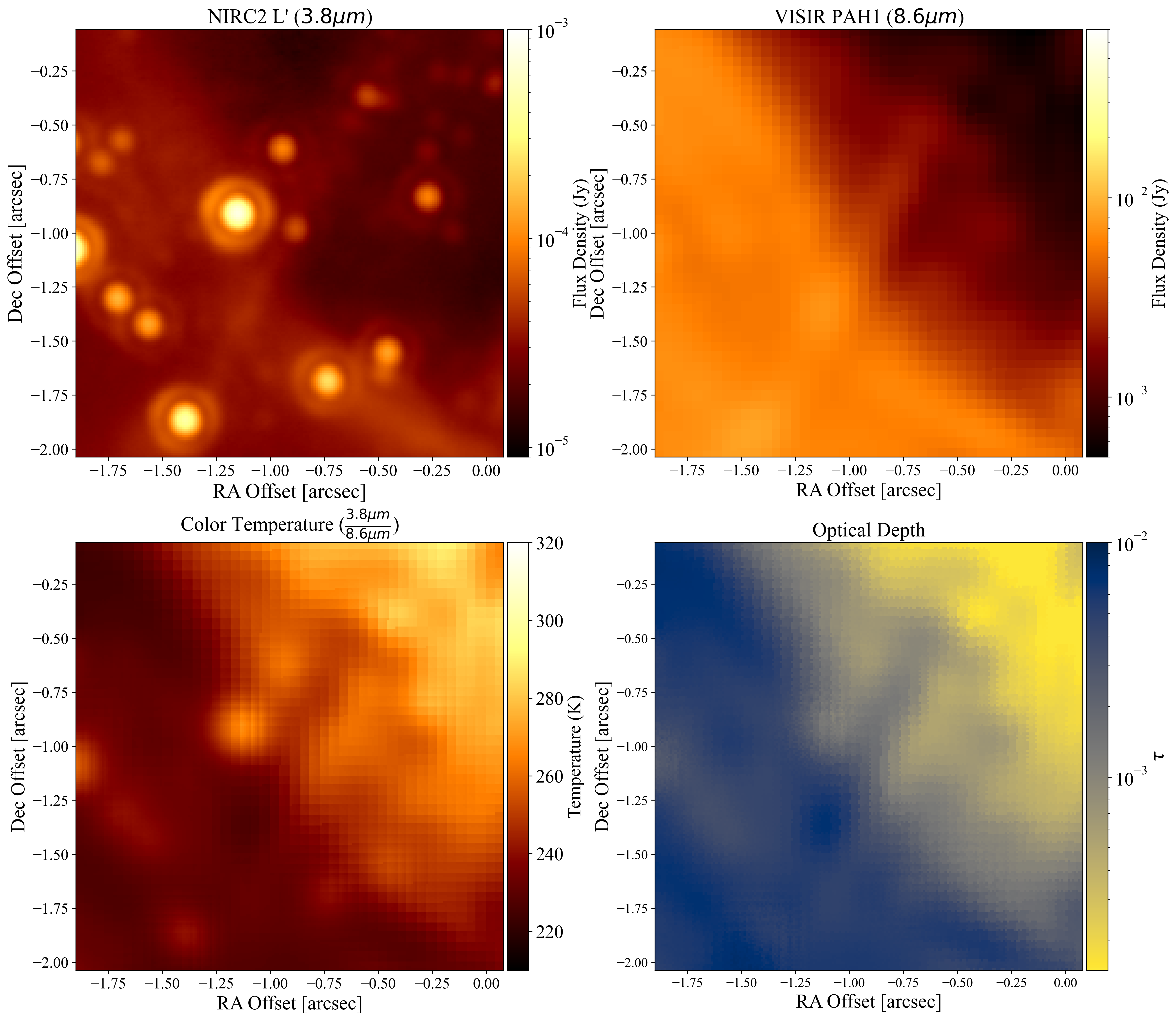}
    \caption{Comparison of the Strand as observed in the 3 3.6~$\mu$m band (upper left), 8.6~$\mu$m band (upper right), color temperature map (lower left), and optical depth map at 3.8~$\mu$m (lower right). The strand is manifested in the color temperature map as a linear depression. The color temperature map indicates the possible presence of a secondary feature (Strand 2), parallel to the Strand}
    \label{fig:linear_comp}
\end{figure}

Bisecting the Northern Arm in projection is a filamentary feature which we  refer to here as the "Strand" (see Figure \ref{fig:linear_comp}); it has previously been described by \citet{Habuois2012A&A...540A..41H} as the "Snake"  and by \citet{Eckart2006A&A...455....1E} as the "Linear Feature". The geometry and proximity to Sgr~A* suggests an outflow nature, for example a jet (\citealt{Eckart2006A&A...455....1E}), although the Strand cannot be followed closer than about 1 arcsecond from Sgr A*, and its northernmost detectable portion points slightly to the eastern side of Sgr A*. The color temperature map indicates the possible presence of a faint secondary component (Strand 2) parallel to this feature. The two components appear as cool features, separated by a warmer gap. 

The color map indicates that the feature's color temperature is comparable to that of the northern arm, which it crosses at an angle of about 60$^\circ$. 

\begin{figure}
    \centering
    \includegraphics[width=0.98\textwidth]{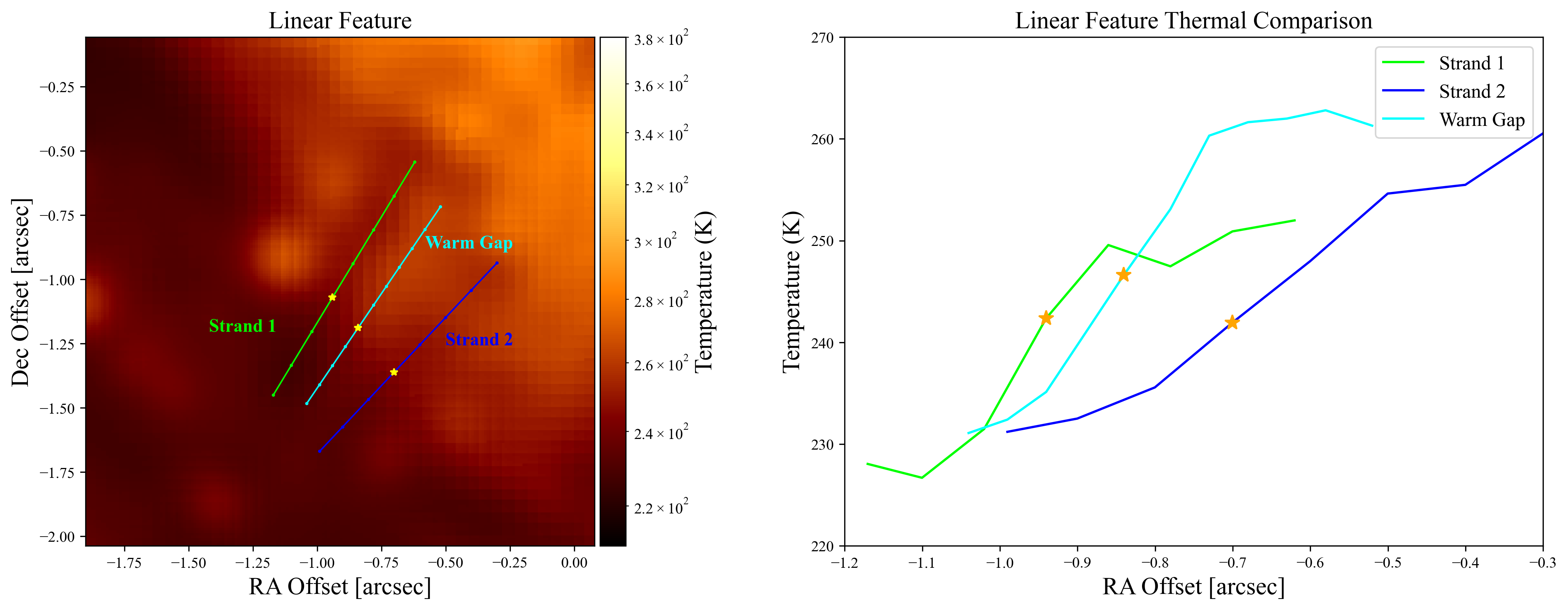}
    \caption{Color temperature trends along the two Strands and the warm gap between them.  Offsets shown on the coordinate axes are relative to Sgr A*.}
    \label{fig:linear_labeled}
\end{figure}

\subsubsection{The Western edge of the Mini-cavity}

\begin{figure}
    \centering
    \includegraphics[width=0.98\textwidth]{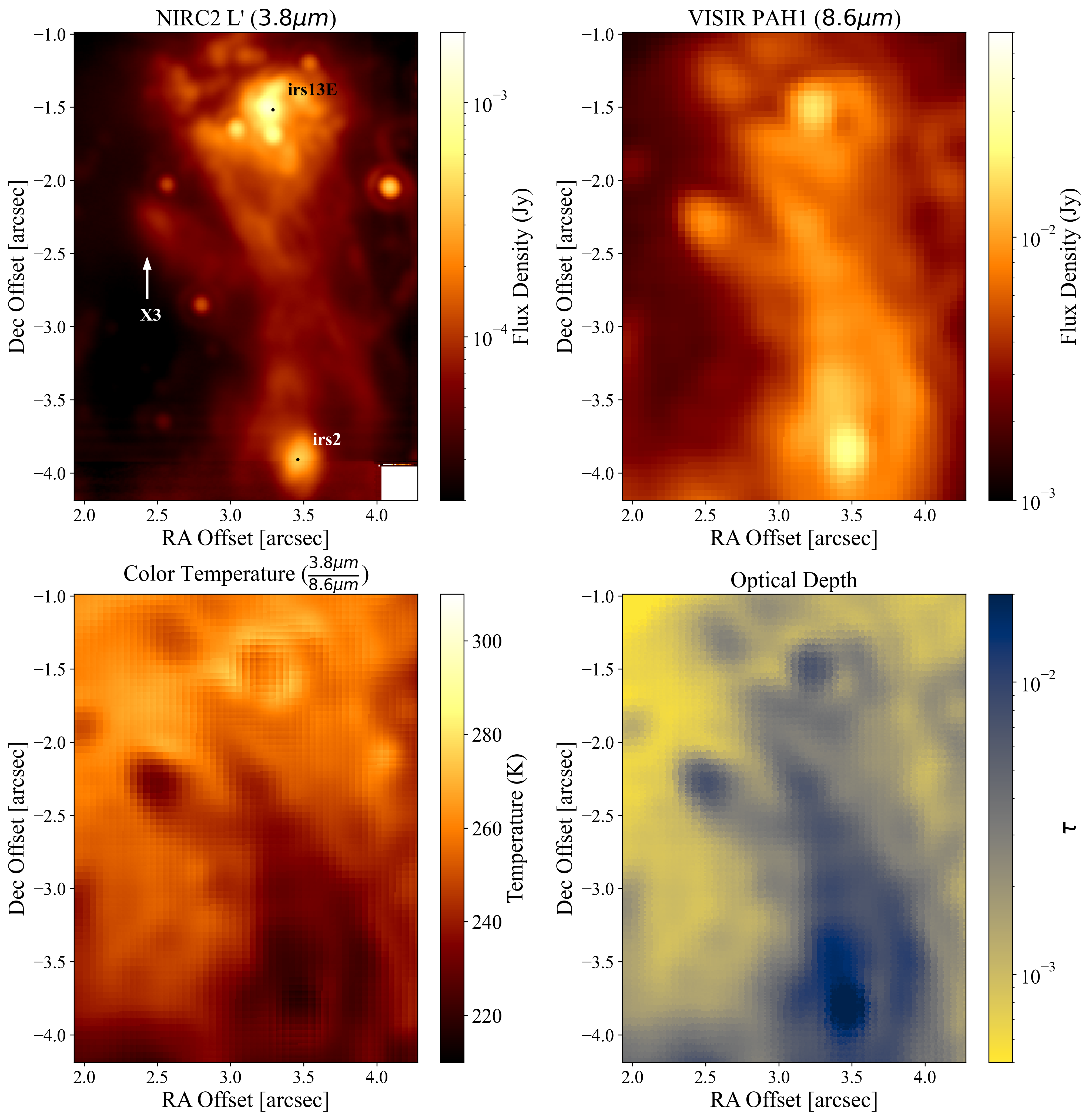}
    \caption{The western edge of the Mini-cavity as observed in the  3.6~$\mu$m band (upper left) , 8.6~$\mu$m band (upper right), color temperature map (lower left), and optical depth map at 3.8~$\mu$m (lower right). The Western Mini-cavity appears to be colder than its surrounding constituents, possibly because of dust shielding of incoming radiation.}
    \label{fig:minicavity}
\end{figure}

The Minicavity is a low-density bubble located in projection at the intersection of the Northern and Eastern Arms of the Sgr~A* West/Mini-spiral. The Mini-cavity's Western edge appears in the color temperature map as a cool structure (Figure\ref{fig:minicavity}), coincident in projection with the cool sources IRS~2 (to the South) and the proposed YSO object X3 (to the East, \citealt{Peissker2023ApJ...944..231P}). While the dust emission appears evenly distributed all the way to the IRS~13E complex in the L' band,  the color temperature, as well as the optical depth map, depict a cooler, crescent-moon shaped feature, defined with a sharp edge resulting from dust shielding incoming radiation. The large column density profile is consistent with previous radio and IR studies. 

%%%%%%%%%%%%%%%%%%%%%%%%%%%%%%%%%%%%%%%%%%%%%%%%%%%%%%%%%%%%%%%%%%%%%%%%%%%%%%%%%
\section{Discussion} \label{sec:disc}
%%%%%%%%%%%%%%%%%%%%%%%%%%%%%%%%%%%%%%%%%%%%%%%%%%%%%%%%%%%%%%%%%%%%%%%%%%%%%%%%%
We note that many of the features identified in this study that have "cool" color temperatures relative to their local backgrounds are associated with relatively bright mid-IR emission, whereas the extended regions having "warm" color temperatures ($\gtrsim280$ K) are systematically rather dim.  This can be understood if the dust in the lowest column density (and thus lowest dust opacity) regions tends to have a higher physical temperature than dust in the highest column density regions.  Indeed, the dust in the lowest column density regions is fully exposed to the high radiation-density environment of the Galactic center and therefore achieves radiative equilibrium at a relatively warm temperature.  On the other hand, although the optical depths that we infer for the "cool" sources are much less than unity at mid-IR wavelengths, these sources have much higher optical depths in the UV and optical, high enough to be able to exclude much of the stellar radiation that would otherwise contribute to heating the dust. As a consequence, we infer that dust in the "cool" color-temperature regions has lower physical temperatures because the dust is largely self-shielded.

Another contributor to the lower temperatures in the regions having the largest optical depths is enhanced cooling in the highest optical depth regions.  Those high column density regions will typically also have the highest densities, and therefore the gas in those regions will have the shortest cooling time.  At gas densities exceeding about $10^4$ cm$^{-3}$, the dust temperature becomes collisionally coupled to that of the gas \citep[c.f.,][eqn 24.6]{Draine_ISM}, so in such high-density regions, the relatively efficient gas cooling can act to cool the dust as well.

%%%%%%%%%%%%%%%%%%%%%%%%%%%%%%%%%%%%%%%%%%%%%%%%%%%%%%%%%%%%%%%%%%%%%%%%%%%%%%%%%
\section{Summary} \label{sec:Summary}
%%%%%%%%%%%%%%%%%%%%%%%%%%%%%%%%%%%%%%%%%%%%%%%%%%%%%%%%%%%%%%%%%%%%%%%%%%%%%%%%%

We have produced the highest-resolution dust-color temperature map and corresponding optical depth map of the inner half parsec of the Galactic Center to date by comparing 3.8~$\mu$m data (in the L' band observed by Keck Observatory's NIRC$2$ imager) and  8.6~$\mu$m data (observed by VLT's VISIR imager in the PAH$1$ band). 

The optical depth map depicts what is essentially an inversion of the temperature map. Lower temperature signatures correspond to higher optical depths.

From the temperature map we have identified 33 hot or cool compact features, measured relative to their local backgrounds. Eleven of these features are newly reported and correspond to candidate dust embedded stellar objects, possibly members of the G-Object population. IRS~13N also appears as a complex of dust-embedded sources: the temperature map reveals three cool clumps associated with the complex. However, while the G objects are isotropically distributed around Sgr~A*, the IRS~13N stars appear closely clustered together. We therefore suggest that the IRS~13N group of stars is simply haphazardly embedded within the very dusty edge of the adjacent Mini-cavity.

We have also characterized several extended features within the region: 
\begin{itemize}
    \item{The Bullet}, a projectile-shaped clump of gas and dust rapidly travelling away from Sgr~A*, has a cold 'head' (49.6~K) tailed by gradually warmer material, possibly formed by a stellar collision. 
    \item{IRS~13E}, a complex of windy stars intermixed with dust and gas, proposed to host an intermediate mass black hole, or alternatively, to contain a shock front that is the product of colliding winds, shows a warm shell surrounding a colder core that coincides with the hypothetical colliding wind source. 
    \item{ID8 and ID11} trace an apparent shock westward of the star IRS~7SE and the dusty neighborhood of the star S4-2 respectively. 
    \item{The Strand}, a filamentary structure bisecting the Northern Arm, appears as a cool feature, possibly associated with a secondary parallel component, separated from the Strand by a warm gap and possibly connected to outflows from Sgr~A*.  
    \item{The Western Edge of the Mini-cavity} appears as a sharp, cool crescent-shaped feature of high column density, likely to be a compressed ridge caused by the expansion of the Mini-cavity. 
\end{itemize}

All these features underline the diversity of phenomena happening in this rich region. Future JWST observations will allow the observation of the dust at many IR wavelengths extending farther into the mid-infrared, which should provide the spectral energy distributions required to generate complete models of the actual dust temperature distributions within the features discussed here. This will help constrain the heating and cooling mechanisms that give rise to the structures that we report here, and ultimately their nature. 

%%%%%%%%%%%%%%%%%%%%%%%%%%%%%%%%%%%%%%%%%%%%%%%%%%%%%%%%%%%%%%%%%%%%%%%%%%%%%%%%%
\begin{acknowledgments}
Support for this work was provided by NSF AAG grant AST-1412615, Jim and Lori Keir, the W. M. Keck Observatory Keck Visiting Scholar program, the Gordon and Betty Moore Foundation, the Heising-Simons Foundation, and Howard and Astrid Preston. A. M. G. acknowledges support from her Lauren B. Leichtman and Arthur E. Levine Endowed Astronomy Chair. 
R. S. acknowledges financial support from the State Agency for Research of the Spanish MCIU through the ''Center of Excellence Severo Ochoa'' award for the Istituto de Astrof\'isica de Andaluc\'ia (SEV-2017-0709) and financial support from national project PGC2018-095049-B- C21 (MCIU/AEI/FEDER, UE).
The W. M. Keck Observatory is operated as a scientific partnership among the California Institute of Technology, the University of California, and the National Aeronautics and Space Administration. The Observatory was made possible by the generous financial support of the W. M. Keck Foundation. 
The authors wish to recognize and acknowledge the very significant cultural role and reverence that the summit of Maunakea has always had within the indigenous Hawaiian community. We are most fortunate to have the opportunity to conduct observations from this mountain.
\end{acknowledgments}

%%%%%%%%%%%%%%%%%%%%%%%%%%%%%%%%%%%%%%%%%%%%%%%%%%%%%%%%%%%%%%%%%%%%%%%%%%%%%%%%%
\pagebreak
\appendix

\section{Compact Sources}\label{sec:sources}

We present here a comprehensive table of sources that stand out from their background in the color temperature map (see Fig. \ref{fig:color_temp}).

\LTcapwidth=\textwidth
\begin{longtable}{lcccccccr}
\caption{Table of sources that stand out from their background.}
\label{tab:VALUES}
\\\hline
NAME & RA ('') & Dec ('') & Lp (mJy) & Ms (mJy) & PAH1 (mJy) & SOURCE (K) & TEMP DIFF (K) & CATEGORY\\
\hline\hline\\
ID1 & -5.9601 & 4.9035 & 35.1$\pm$0.8 &  & 23.4$\pm$1.1 & 209$\pm$3 & -0.1$\pm$0.0 & \\
ID2 & -4.6654 & 5.2794 & $<$0.07 &  & 11.2$\pm$0.5 & 224$\pm$3 & 12.6$\pm$0.4 & hot\\
S6-93* & -4.5458 & 4.9696 & 9.17$\pm$0.2 &  & 12.0$\pm$0.6 & 231$\pm$3 & 7.3$\pm$0.2 & \\
ID4 & -4.3503 & 5.2687 & 20.2$\pm$0.4 &  & 6.24$\pm$0.29 & 251$\pm$4 & 16.0$\pm$0.6 & hot\\
S4-364* & -2.3465 & 4.4677 & 16.9$\pm$0.4 &  & 8.31$\pm$0.39 & 323$\pm$7 & 16.4$\pm$0.8 & hot\\
S6-89* & -5.5158 & 2.9379 & 13.7$\pm$0.3 &  & 11.1$\pm$0.5 & 231$\pm$3 & 11.3$\pm$0.4 & hot\\
ID7 & -3.7925 & 3.048 & 2.0$\pm$0.04 & $<$0.114 & 278$\pm$13 & 219$\pm$3 & -8.3$\pm$0.3 & \\
ID8 & -3.2352 & 3.3777 & 10.0$\pm$0.2 & 20.5$\pm$1.1 & 163$\pm$7 & 230$\pm$3 & -5.1$\pm$0.2 & \\
ID9 & -5.7845 & 2.1271 & 9.92$\pm$0.22 &  & 11.3$\pm$0.5 & 229$\pm$3 & 8.3$\pm$0.3 & \\
IRS1W & -5.2449 & 0.6479 & 1137$\pm$25 &  & 16885$\pm$795 & 223$\pm$3 & 3.9$\pm$0.1 & \\
ID11 & -3.5415 & 1.5572 & 14.1$\pm$0.3 & 34.2$\pm$1.8 & 636$\pm$29 & 219$\pm$3 & -9.0$\pm$0.3 & \\
IRS16NE & -2.9322 & 0.8705 & 218$\pm$4 & 165$\pm$8 & 276$\pm$13 & 280$\pm$5 & 42.6$\pm$1.7 & hot\\
IRS16CC & -1.9563 & 0.6352 & 50.0$\pm$1.1 & 40.6$\pm$2.1 & 82.4$\pm$3.9 & 259$\pm$4 & 16.1$\pm$0.6 & hot\\
MIR Cometary  & -1.3159 & 0.9394 & 1.52$\pm$0.03 & 5.79$\pm$0.3 & 144$\pm$6 & 261$\pm$5 & -27.7$\pm$1.2 & cool\\
IRS16C & -0.9573 & 0.5893 & 113$\pm$2 & 97.9$\pm$5.0 & 150$\pm$7 & 322$\pm$7 & 24.1$\pm$1.1 & hot\\
IRS16NW & -0.0863 & 1.2434 & 89.2$\pm$2.0 & 65.2$\pm$3.4 & 116$\pm$5 & 335$\pm$8 & 7.6$\pm$0.4 & \\
ID17 & -0.0054 & 2.5863 & $<$0.04 & $<$0.079 & 17.7$\pm$0.8 & 318$\pm$7 & -22.9$\pm$1.2 & cool\\
ID18 & 0.3662 & 2.2579 & $<$0.04 & $<$0.076 & 7.42$\pm$0.35 & 332$\pm$8 & -5.4$\pm$0.3 & \\
ID19 & 0.8067 & 2.0344 & 2.29$\pm$0.05 & $<$0.079 & 10.9$\pm$0.5 & 342$\pm$8 & 1.4$\pm$0.1 & \\
S2-16 & 1.1629 & 2.054 & 52.7$\pm$1.2 & 56.8$\pm$2.9 & 119$\pm$5 & 333$\pm$8 & -15.0$\pm$0.8 & cool\\
IRS29N & 1.5393 & 1.3583 & 538$\pm$11 & 693$\pm$35 & 1895$\pm$89 & 299$\pm$6 & -23.4$\pm$1.1 & cool\\
IRS3E & 2.3409 & 3.8371 & 1143$\pm$25 &  & 82185$\pm$3872 & 208$\pm$3 & -22.0$\pm$0.7 & cool\\
bullet & 3.6159 & 2.2442 & 2.97$\pm$0.07 & 7.19$\pm$0.37 & 393$\pm$18 & 234$\pm$4 & -46.1$\pm$1.8 & cool\\
ID24 & 3.8309 & 1.9064 & 34.5$\pm$0.8 & 43.9$\pm$2.3 & 247$\pm$11 & 263$\pm$5 & -2.5$\pm$0.1 & \\
ID25 & 3.7457 & 1.453 & 14.2$\pm$0.3 & 33.3$\pm$1.7 & 137$\pm$6 & 266$\pm$5 & -12.2$\pm$0.5 & cool\\
IRS34W* & 4.1038 & 1.5057 & 59.8$\pm$1.3 & 61.6$\pm$3.2 & 38.7$\pm$1.8 & 293$\pm$6 & 29.8$\pm$1.3 & hot\\
S3-10 & -3.3523 & -1.073 & 10.3$\pm$0.2 & 6.44$\pm$0.33 & 17.4$\pm$0.8 & 227$\pm$3 & 7.1$\pm$0.2 & \\
S3-34 & -3.232 & -1.3113 & 4.11$\pm$0.09 & 3.71$\pm$0.19 & 16.7$\pm$0.8 & 226$\pm$3 & 3.8$\pm$0.1 & \\
S3-5 & -2.9935 & -1.123 & 22.5$\pm$0.5 & 18.8$\pm$1.0 & 16.8$\pm$0.8 & 236$\pm$4 & 10.2$\pm$0.4 & hot\\
IRS16SW-E & -1.9328 & -1.1025 & 106$\pm$2 & 97.6$\pm$5.0 & 229$\pm$10 & 256$\pm$4 & 26.2$\pm$1.0 & hot\\
S2-6 & -1.6958 & -1.3489 & 20.7$\pm$0.5 & 12.4$\pm$0.6 & 17.3$\pm$0.8 & 239$\pm$4 & 7.5$\pm$0.3 & \\
S2-4 & -1.5662 & -1.4293 & 16.4$\pm$0.4 & 13.7$\pm$0.7 & 16.9$\pm$0.8 & 238$\pm$4 & 7.3$\pm$0.3 & \\
ID33 & -1.1122 & -1.3688 & 4.16$\pm$0.09 & 12.1$\pm$0.6 & 306$\pm$14 & 227$\pm$3 & -6.2$\pm$0.2 & \\
IRS16SW & -1.1384 & -0.9192 & 91.4$\pm$2.0 & 85.9$\pm$4.4 & 13.8$\pm$0.6 & 264$\pm$5 & 28.4$\pm$1.1 & hot\\
ID35 & -0.4693 & -0.3657 & 4.92$\pm$0.11 & $<$0.063 & 2.94$\pm$0.14 & 278$\pm$5 & 6.1$\pm$0.3 & \\
X7 & 0.5588 & -0.343 & 7.44$\pm$0.17 & 14.4$\pm$0.7 & 91.6$\pm$4.3 & 262$\pm$5 & -12.2$\pm$0.5 & cool\\
epsilon source & 1.6474 & -0.4435 & 10.6$\pm$0.2 & 18.4$\pm$0.9 & 312$\pm$14 & 235$\pm$4 & -17.9$\pm$0.7 & cool\\
ID38 & 1.9505 & -0.1366 & 0.61$\pm$0.01 & 3.95$\pm$0.2 & 155$\pm$7 & 251$\pm$4 & -17.4$\pm$0.7 & cool\\
ID39 & 2.8933 & -0.7437 & 23.5$\pm$0.5 & 48.5$\pm$2.5 & 305$\pm$14 & 247$\pm$4 & -9.3$\pm$0.4 & \\
ID40 & 4.265 & 0.3066 & $<$0.05 & $<$0.047 & 4.01$\pm$0.19 & 246$\pm$4 & 3.0$\pm$0.1 & \\
S4-129 & -3.7352 & -2.2387 & 13.4$\pm$0.3 & 5.04$\pm$0.26 & 7.49$\pm$0.35 & 248$\pm$4 & 11.9$\pm$0.4 & hot\\
S4-4 & -3.6049 & -2.5178 & 11.8$\pm$0.3 & 5.0$\pm$0.26 & 5.89$\pm$0.28 & 255$\pm$4 & 7.8$\pm$0.3 & \\
S4-6 & -3.3064 & -2.8163 & 8.18$\pm$0.18 & 3.37$\pm$0.17 & 6.09$\pm$0.29 & 258$\pm$4 & 4.8$\pm$0.2 & \\
IRS21 & -2.3264 & -2.7246 & 859$\pm$19 & 1150$\pm$59 & 14539$\pm$685 & 231$\pm$3 & -8.2$\pm$0.3 & \\
S2-17 & -1.4055 & -1.8917 & 52.5$\pm$1.2 & 39.2$\pm$2.0 & 371$\pm$17 & 239$\pm$4 & 10.0$\pm$0.3 & \\
S1-24 & -0.7614 & -1.718 & 27.9$\pm$0.6 & 30.4$\pm$1.6 & 17.9$\pm$0.8 & 237$\pm$4 & 6.0$\pm$0.2 & \\
S1-17 & -0.4411 & -1.5649 & 19.6$\pm$0.4 & 13.3$\pm$0.7 & 8.71$\pm$0.41 & 251$\pm$4 & 5.8$\pm$0.2 & \\
ID48 & 0.8743 & -1.3375 & 17.0$\pm$0.4 & 20.6$\pm$1.1 & 194$\pm$9 & 237$\pm$4 & -12.4$\pm$0.5 & cool\\
IRS33N & 0.0054 & -2.2888 & 27.8$\pm$0.6 & 19.5$\pm$1.0 & 16.0$\pm$0.8 & 242$\pm$4 & 13.7$\pm$0.5 & hot\\
S1-23 & 0.8871 & -1.5308 & 25.4$\pm$0.6 & 23.6$\pm$1.2 & 115$\pm$5 & 248$\pm$4 & 3.4$\pm$0.1 & \\
ID51 & 0.4483 & -2.26 & 9.42$\pm$0.21 & 29.0$\pm$1.5 & 602$\pm$28 & 221$\pm$3 & -13.2$\pm$0.5 & cool\\
ID52 & 0.8208 & -2.3416 & $<$0.07 & $<$0.099 & 70.4$\pm$3.3 & 243$\pm$4 & -0.7$\pm$0.0 & \\
ID53 & 1.4471 & -2.081 & $<$0.06 & $<$0.063 & 69.4$\pm$3.3 & 251$\pm$4 & -4.5$\pm$0.2 & \\
ID54 & 1.9559 & -1.9213 & 1.8$\pm$0.04 & 5.13$\pm$0.26 & 234$\pm$11 & 250$\pm$4 & -11.1$\pm$0.4 & cool\\
ID55 & 2.574 & -1.2637 & 23.6$\pm$0.5 & 53.5$\pm$2.8 & 451$\pm$21 & 250$\pm$4 & -10.9$\pm$0.4 & cool\\
X3 & 2.4927 & -2.288 & 21.7$\pm$0.5 & 51.1$\pm$2.6 & 941$\pm$44 & 232$\pm$3 & -23.3$\pm$0.8 & cool\\
IRS13E & 3.1948 & -1.5099 & 281$\pm$6 & 414$\pm$21 & 1572$\pm$74 & 258$\pm$4 & -5.9$\pm$0.2 & \\
ID58 & -5.5013 & -4.2489 & $<$0.03 &  & 70.7$\pm$3.3 & 234$\pm$4 & -20.3$\pm$0.7 & cool\\
ID59 & -5.0244 & -3.7006 & $<$0.04 & $<$0.07 & 2.59$\pm$0.12 & 258$\pm$4 & 3.8$\pm$0.1 & \\
S5-183 & -4.5859 & -3.4734 & 18.7$\pm$0.4 & 11.0$\pm$0.6 & 2.77$\pm$0.13 & 275$\pm$5 & 21.5$\pm$0.9 & hot\\
IRS33E & -0.7356 & -3.1634 & 89.2$\pm$2.0 & 70.1$\pm$3.6 & 207$\pm$9 & 258$\pm$4 & 25.3$\pm$0.9 & hot\\
S4-71 & -0.7557 & -4.1432 & 11.2$\pm$0.2 & 6.04$\pm$0.31 & 7.44$\pm$0.35 & 237$\pm$4 & 8.8$\pm$0.3 & \\
S3-22 & 0.3267 & -3.2222 & 37.1$\pm$0.8 & 22.8$\pm$1.2 & 8.6$\pm$0.41 & 256$\pm$4 & 15.9$\pm$0.6 & hot\\
ID64 & 0.5932 & -3.5785 & 2.02$\pm$0.04 & 3.93$\pm$0.2 & 221$\pm$10 & 232$\pm$4 & -8.1$\pm$0.3 & \\
ID65 & 1.3641 & -3.1802 & 2.87$\pm$0.06 & 9.39$\pm$0.48 & 167$\pm$7 & 233$\pm$4 & -10.8$\pm$0.4 & cool\\
ID66 & 1.8765 & -3.2002 & 3.86$\pm$0.09 & 11.1$\pm$0.6 & 122$\pm$5 & 238$\pm$4 & -7.5$\pm$0.3 & \\
ID67 & 2.8152 & -4.2098 & $<$0.11 & $<$0.196 & 13.1$\pm$0.6 & 233$\pm$4 & 2.5$\pm$0.1 & \\
IRS2 & 3.4539 & -3.8028 & 108$\pm$2 & 198$\pm$10 & 2738$\pm$129 & 220$\pm$3 & -6.5$\pm$0.2 & \\
\hline\hline\\
\end{longtable}
\vspace{-4.5em}
\begin{longtable}{l}
  \hspace*{6mm}For each source, we report the position as an offset from Sgr~A*, the flux at 3.8 and 8.6~$\mu$m, the temperature, the temperature\\
  difference with respect to the local background and classify them as hot/cool if they are different enough from their background.\\ Sources were identified with the catalogue reported in \cite{Gautam2019ApJ...871..103G}, with the exception of sources denoted with a (*),\\
  which were identified with the catalogue reported in \citep{Yelda2014ApJ...783..131Y}. Newly reported sources have been denoted with an 'ID' \\ tag. ID~38 correlates to Source B in \cite{Yusef_Zadeh_2016ApJ...819...60Y}. \\
  \hline\\
\end{longtable}

\bibliography{main}{}
\bibliographystyle{aasjournal}



\end{document}